\documentclass[twocolumn]{aastex631}
\usepackage{silence}
\usepackage{graphicx}
\usepackage[caption=false]{subfig}
\usepackage{booktabs}
\usepackage{stmaryrd}
\usepackage{xcolor}
\usepackage{natbib}
\setcitestyle{authoryear,open={(},close={)}}

\usepackage{float}
\usepackage{color,soul}
\usepackage{wrapfig}
\usepackage{bm}
\usepackage{amsmath}

\usepackage{fancyhdr}
\usepackage{wasysym}

\newcommand{\lir}{$L_{\mathrm{IR}}$}
\newcommand{\loglir}{$\log\mathrm{L}_{\mathrm{IR}}/\mathrm{L}_\odot$}

\begin{document}

\title{A JWST MIRI LRS Survey of 37 Massive Star-forming Galaxies and Active Galactic Nuclei at Cosmic Noon: Overview and First Results}
\author[0000-0002-6149-8178]{Jed McKinney}
\altaffiliation{NASA Hubble Fellow}
\affiliation{Department of Astronomy, The University of Texas at Austin, Austin, TX, USA}
\affiliation{Cosmic Frontier Center, The University of Texas at Austin, Austin, TX 78712}

\author[0000-0002-4690-4502]{Miriam Eleazer}
\affiliation{Department of Astronomy, University of Massachusetts, Amherst, MA 01003, USA.}

\author[0000-0001-8592-2706]{Alexandra Pope} 
\affiliation{Department of Astronomy, University of Massachusetts, Amherst, MA 01003, USA.}

\author[0000-0002-1917-1200]{Anna Sajina}
\affiliation{Department of Physics and Astronomy, Tufts University, Medford, MA 02155, USA}

\author[0000-0002-8909-8782]{Stacey Alberts}
\affiliation{AURA for the European Space Agency (ESA), Space Telescope Science Institute, 3700 San Martin Dr., Baltimore, MD 21218, USA}

\author[0000-0002-9720-3255]{Meredith Stone}
\affiliation{Steward Observatory, University of Arizona, 933 North Cherry Avenue, Tucson, AZ 85719, USA}

\author[0000-0002-8502-8947]{Leonid Sajkov}
\affiliation{Department of Physics and Astronomy, Tufts University, Medford, MA 02155, USA}

\author[0009-0001-4693-1519]{Virginia Vanicek}
\affiliation{Department of Astronomy, The University of Texas at Austin, Austin, TX, USA}

\author[0000-0002-5537-8110]{Allison Kirkpatrick}
\affiliation{Department of Physics \& Astronomy, University of Kansas, Lawrence, KS 66045, USA}

\author[0000-0001-8490-6632]{Thomas S.-Y. Lai}
\affiliation{IPAC, California Institute of Technology, 1200 E. California Blvd., Pasadena, CA 91125, USA}

\author[0000-0002-0930-6466]{Caitlin M. Casey}
\affiliation{Department of Physics, University of California Santa Barbara, Santa Barbara, CA, USA}

\author[0000-0003-3498-2973]{Lee Armus}
\affiliation{IPAC, California Institute of Technology, 1200 E. California Blvd., Pasadena, CA 91125, USA}

\author[0000-0003-0699-6083]{Tanio D\'{i}az-Santos}
\affiliation{Institute of Astrophysics, Foundation for Research and Technology-Hellas (FORTH), Heraklion, GR-70013, Greece}
\affiliation{School of Sciences, European University Cyprus, Diogenes street, Engomi, 1516 Nicosia, Cyprus}

\author{Andrew Korkus}
\affiliation{Department of Physics and Astronomy, Tufts University, Medford, MA 02155, USA}

\author[0000-0003-3881-1397]{Olivia Cooper}
\affiliation{Department of Astronomy, The University of Texas at Austin, Austin, TX, USA}

\author[0000-0002-1496-6514]{Lindsay R. House}
\affiliation{Department of Astronomy, The University of Texas at Austin, Austin, TX, USA}

\author[0000-0003-3596-8794]{Hollis Akins}\altaffiliation{NSF Graduate Research Fellow}
\affiliation{Department of Astronomy, The University of Texas at Austin, Austin, TX, USA}

\author[0000-0003-3216-7190]{Erini Lambrides}\altaffiliation{NPP Fellow}
\affiliation{NASA-Goddard Space Flight Center, Code 662, Greenbelt, MD, 20771, USA}

\author[0000-0002-7530-8857]{Arianna S. Long}
\affiliation{Department of Astronomy, University of Washington, Seattle, WA, USA}

\author[0000-0003-1710-9339]{Lin Yan}
\affiliation{The Caltech Optical Observatories, California Institute of Technology, Pasadena, CA 91125, USA}

\begin{abstract}
We present a large spectroscopic survey with \textit{JWST}'s Mid-Infrared Instrument Low Resolution Spectrometer (LRS) targeting $37$ infrared-bright galaxies between $z=0.65-2.46$ with infrared luminosities $\log L_{\rm IR}/L_\odot>11.5$ and $\log M_*/M_\odot=10-11.5$.
Targets were taken from a \textit{Spitzer} $24\,\mu$m selected sample with archival spectroscopy from the Infrared Spectrograph (IRS) and include a mix of star-forming galaxies and dust-obscured active galactic nuclei. By combining IRS with the increased sensitivity of LRS, we expand the range of spectral features observed between $5$ and $30\,\mu$m for every galaxy in our sample. In this paper, we outline the sample selection, reduction of the \textit{JWST} data, extraction of the 1D spectra, and 
polycyclic aromatic hydrocarbon (PAH) feature measurements from $\lambda_{\rm rest}=3.3-11.3\,\mu$m. In the \textit{JWST} spectra, we detect PAH emission features at $3.3-5.3\,\mu$m, as well as Paschen and Brackett lines. The $3.3\,\mu$m feature can be as bright as $0.2\%$ of the $8-1000\,\mu$m infrared luminosity and exhibits a tight correlation with the dust-obscured star formation rate. We detect absorption features from CO gas, CO$_2$ ice, H$_2$O ice, and aliphatic dust. From the joint \textit{JWST} and \textit{Spitzer} analysis, we find that the $11.3/3.3\,\mu$m PAH ratios are on average three times higher than that of local luminous infrared galaxies. This is interpreted as 
evidence that the PAH grains are larger at $z\sim1-2$.  
The size distribution may be affected by coagulation of grains due to high gas densities and low temperatures. These conditions are supported by the observation of strong water ice absorption at $3.05\,\mu$m, and can lower stellar radiative feedback because large PAHs transmit less energy per photon into the interstellar medium.
\end{abstract}

\keywords{Luminous infrared galaxies ---Ultraluminous infrared galaxies --- Galaxy evolution --- Infrared spectroscopy}

\section{Introduction\label{sec:intro}}
The majority of star formation over the last $10$ Gyr is obscured by dust within the most luminous infrared galaxies \citep{Gruppioni2020,Zavala2021,Traina2024}. The relationship between star formation and dust is complex, with the properties of the grains themselves playing a fundamental role in regulating gas cooling, heating, shielding, and molecular hydrogen formation \citep{Watson1972,Bakes1994,Helou2001,Soliman2024}. Dust is therefore not just an important by-product of stellar evolution but also an active player in galaxy formation by mediating interstellar medium (ISM) conditions and facilitating future star formation. 

It is commonly assumed that the properties of dust in distant galaxies, such as grain size distributions, charge, and composition, are analogous to the dust along lines of sight in the Milky Way or in nearby galaxies where these properties are well constrained \citep[e.g.,][]{WeingartnerDraine2001,Galliano2018}. While this assumption is appropriate and well motivated for interpreting galaxy-integrated dust, such as thermal emission of cold grains and general extinction properties, it may fail on star formation region scales where dust is actively regulating stellar feedback \citep{Krumholz2011,Glover2012,Soliman2024}. In nearby galaxies, the properties of dust grains are observed to strongly correlate with properties of the interstellar radiation field (ISRF), such as its density and hardness \citep[e.g.,][]{Rigopoulou2021,Draine2021pah,Baron2025}. High-redshift galaxies exhibit larger ionization parameters\footnote{The ionization parameter is defined as the ratio of the average hydrogen-ionizing photon flux to the hydrogen density.} than local galaxies \citep{Kaasinen2018,Shen2025}, and the cold gas conditions exhibit higher excitation conditions \citep{Boogaard2020}.  High-redshift galaxies also have on average larger total dust masses and warmer dust temperatures \citep{Bethermin2015,Schreiber2018,Ismail2023,Koprowski2024,Hill2025}. Therefore, we may also expect grain properties such as size distributions and ionization states to differ from those commonly found in the nearby Universe. Given that grain properties factor into important astrophysical processes such as neutral gas heating \citep[e.g.,][]{Bakes1994,WeingartnerDraine2001}, the impact of these differences may be important for instantaneous and prolonged star formation in high-redshift galaxies. 



The emission profile of dust across the electromagnetic spectrum provides numerous ways to assess quantities such as the shape, size, charge, and composition of grains (see e.g., \citealt{Hensley2023}). Polycyclic aromatic hydrocarbons (PAHs) are the most common dust grain by number at sizes $a<0.005\,\mu$m ($\approx50-100$ C atoms) and are thought to be responsible for the luminous broad features observed at near- through mid-infrared wavelengths \citep[e.g.,][]{Mattila1996,Verstraete1996,Smith2007,Sajina2007,Stierwalt2013}. 
These PAH bands encode the grains' underlying size and charge characteristics \citep{Draine2001,Peeters2004,Draine2007,Maragkoudakis2020}, with ionized grains having stronger $6.2$, $7.7$, and $8.3\,\mu$m features as compared to the neutral grains having stronger features at $3.3$ and $11.3\,\mu$m. Despite comprising $\lesssim5\%$ of the total dust mass \citep{Draine2007}, PAHs are responsible for the dominant ISM heating channel (via photoelectrons) which contributes to the multiphase structure and evolution of gas \citep{Bakes1994,Hollenbach1999}. 



PAHs span the near- to mid-infrared with the strongest features at $3.3$, $6.2$, $7.7$, $8.6$, $11.3$, and $17\,\mu$m. Individually, these features and the underlying continuum have been used as active galactic nuclei (AGN) indicators and to trace star formation rates \citep[SFRs; e.g.,][]{Spoon2007,Calzetti2007,Kim2012,Shipley2016,Inami2018}; however, the power of PAHs as diagnostics on the grain size distribution, ionization, and incident radiation field strengths is accessed by combinations of feature ratios \citep{Draine2007,Maragkoudakis2020}. 
When combined with longer wavelength features, the $3.3\,\mu$m PAH adds the most diagnostic power on the PAH grain size distribution \citep{Maragkoudakis2020}, which plays an important role in setting heating/cooling rates in the ISM \citep{Tielens1985}. The $3.3\,\mu$m PAH is also the least well studied: inaccessible by the \textit{Spitzer Space Telescope}'s Infrared Spectrograph (IRS) at $z=0$ and too faint to detect with \textit{Spitzer} in most distant galaxies. \cite{Sajina2009} presented the first detection of the $3.3\,\mu$m line in a $z\sim2$ ultraluminous infrared galaxy (ULIRG, $L_{\rm IR}/L_\odot>10^{12}$), which was quickly followed by a detection of this feature in an even brighter gravitionally lensed ULIRG at $z=3.075$ by \cite{Siana2009}. Between then and the launch of \textit{JWST}, the $3.3\,\mu$m PAH was only detected in four galaxies beyond $z\sim0$. 

\cite{Inami2018} used \textit{AKARI} to present the first statistical analysis of the $3.3\,\mu$m PAH feature targeting a subset of local luminous infrared galaxies selected from the Great Observatories All Sky LIRG Survey (GOALS, \citealt{Armus2007}). \cite{Lai2020} presented a comprehensive analysis of all PAH features between $3.3$ and $17\,\mu$m in 113 star-forming galaxies based on combined \textit{AKARI} and \textit{Spitzer} spectra. \cite{Inami2018} found the $3.3\,\mu$m PAH to be bright and a good indicator of the relative balance between star formation and supermassive black hole accretion. Using these same data, \cite{McKinney2021a} demonstrated that important changes in the photoelectric heating efficiency of local luminous infrared galaxies (LIRGs, $L_{\rm IR}/L_\odot>10^{11})$ can only be teased out when folding the $3.3\,\mu$m PAH line into joint analysis with the longer wavelength PAH features and far-infrared fine-structure lines that encode gas cooling rates. 

With the advent of \textit{JWST}, progress is being made observing PAHs, particularly using the $3.3\,\mu$m feature. \cite{Lyu2025} used the Wide-field Slitless Spectrometer on NIRCam to measure $3.3\,\mu$m PAH emission in $88$ galaxies at $z=0.2-0.45$. \cite{Spilker2023} reported a detection in a lensed submillimeter-luminous galaxy at $z\sim4$, making this the highest redshift detection of PAHs in emission to-date. In principle, \textit{JWST} is capable of measuring the $3.3\,\mu$m PAH out to $z\sim4.5$, making this an attractive feature for probing dust and obscured star formation into the early Universe; however, no large survey of this feature has calibrated the relationship between the $3.3\,\mu$m feature and SFR versus AGN diagnostics beyond $z=0.5$. Programs with the \textit{JWST} Mid-Infrared Instrument (MIRI) Medium Resolution Spectrometer (MRS) are capable of obtaining $5-28\,\mu$m spectra and have had great success studying the mid-infrared spectrum of $z>0.5$ LIRGs \citep{Young2023,Sajina2025}; however, the substantial MIRI/MRS gains in spatial and spectral resolving power come at the cost of longer exposure requirements, ultimately yielding smaller samples.  

In this work, we present the first results from a \textit{JWST}/MIRI Low Resolution Spectrometer (LRS) survey of $z\sim1-2$ galaxies targeting the $3.3\,\mu$m PAH and adjacent features at rest-frame $\sim2-\,5\,\mu$m (observed-frame $\sim5-14\,\mu$m). 
Section \ref{sec:sample} presents the selection of our targets, which all have \textit{Spitzer} data that extend the spectral coverage of each galaxy up to rest-frame $\sim12\,\mu$m (observed-frame $\sim30\,\mu$m). The \textit{JWST} observations are described in Section \ref{sec:obs}. In Section \ref{sec:data}, we present our reduction of the MIRI data and the steps taken to produce our final 1D spectra. We describe our analysis of these spectra in Section \ref{sec:analysis} and present the results in Section \ref{sec:results}. Section \ref{sec:disc} discusses the implications of these findings, which we summarize in our conclusion in Section \ref{sec:conc}. Throughout this work, we assume a $\Lambda$CDM cosmology with $H_0=70$\,km\,s$^{-1}$\,Mpc$^{-1}$, $\Omega_m=0.3$, $\Omega_\Lambda=0.7$. 


\section{Sample Selection\label{sec:sample}}

The goal of this work is to build a spectroscopic sample of $z\gtrsim1$ galaxies with detections of PAHs between $3.3$ and $17\,\mu$m. 
Our targets for \textit{JWST} observations were selected from a parent sample of 343 LIRGs with archival \textit{Spitzer}/IRS spectra across the Great Observatories Origins Deep Survey North/South (GOODS-N/S) and the Extended Chandra First Look Survey (xFLS). These galaxies have infrared luminosities (\lir\ integrated between $8$ and $1000\,\mu$m) measured from \textit{Herschel Space Observatory} observations. For a thorough review of the parent sample, see \citet{Kirkpatrick2015}, which documents the compilation of individual IRS programs from \citet{Yan2007,Pope2008,Dasyra2009,Fadda2010,Pope2013}. 
The parent sample was selected to be mid-infrared--bright with \textit{Spitzer}/MIPS 24$\,\mu$m fluxes $\gtrsim100\,\mu$Jy and is broadly representative of both \textit{Spitzer}- and \textit{Herschel}-selected sources \citep{Kirkpatrick2012}. The majority ($\sim80\%$) of these sources fall above the star formation main sequence \citep{Kirkpatrick2017} where all local LIRGs are found \citep{Armus2007,U2012}.  

Achieving a relatively large sample by leveraging archival \textit{Spitzer} spectra was a major goal of this proposal. From the parent sample, we first select candidate targets with spectroscopic redshifts derived from the archival \textit{Spitzer} spectra\footnote{Note that such redshifts are typically accurate to $\Delta z\approx0.01-0.1$ because they are based on broad PAH features as opposed to narrow atomic lines \citep{McKinney2020}.} between $0.7<z<2.2$ such that the $3.3\,\mu$m PAH feature would fall within the $5-14\,\mu$m wavelength range of MIRI/LRS, bringing the number of candidates down from 343 sources to 157. We predict $3.3\,\mu$m PAH fluxes based on conservative scaling relations derived from joint \textit{AKARI} and \textit{Spitzer} observations made of $z\sim0$ LIRGs from \cite{Stierwalt2013,Stierwalt2014} and \cite{Inami2018}. These predictions take into account the IRAC $8\,\mu$m photometry of our candidate targets, which provides known fluxes in the LRS wavelength range. We then require that the predicted $3.3\,\mu$m PAH flux be detectable in $\lesssim1$ hr of exposure time with LRS. 

\begin{figure}
    \includegraphics[width=.48\textwidth]{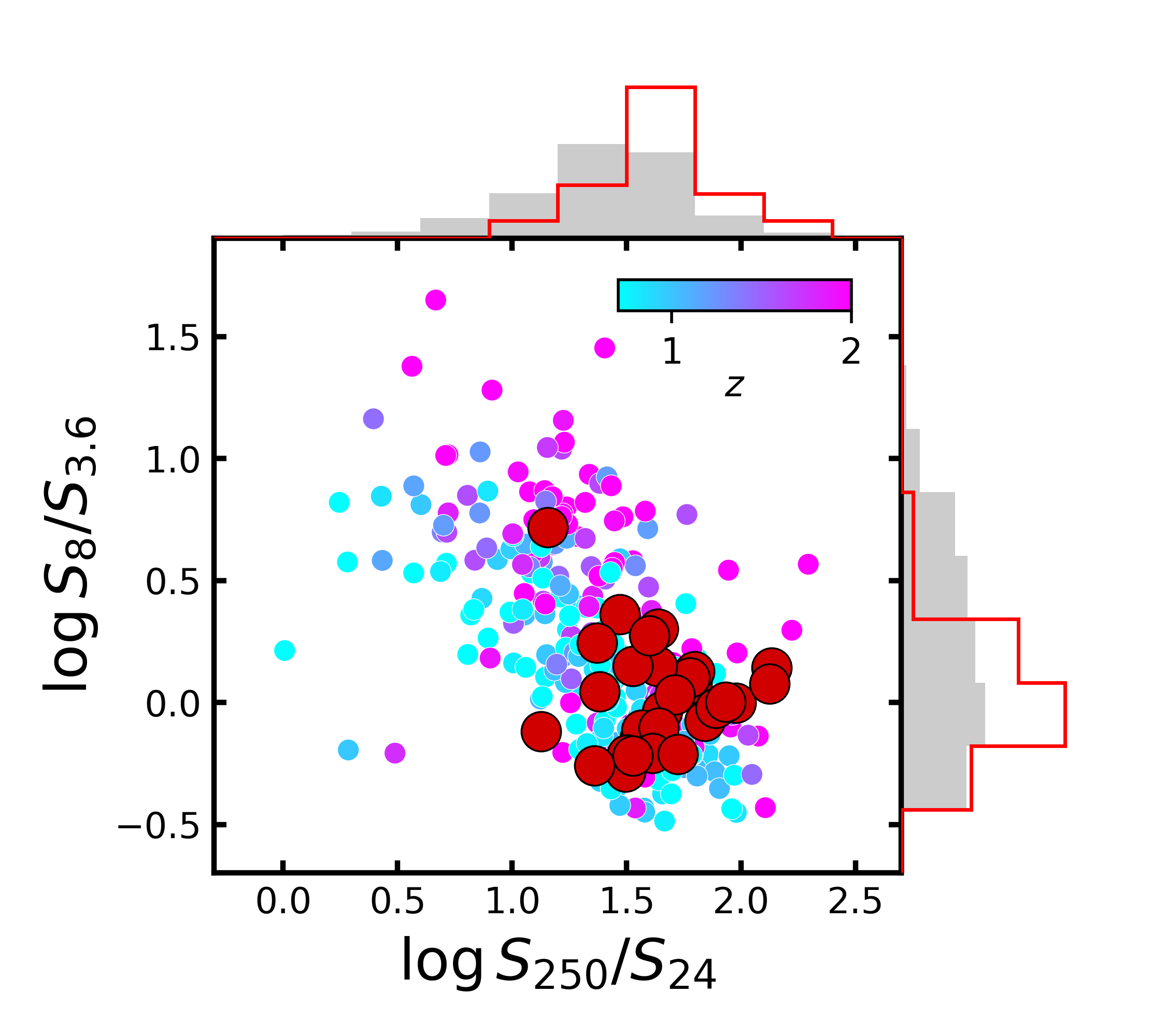}
    \caption{Observed 8$\,\mu$m/$3.6\,\mu$m ratio to 250$\,\mu$m/$24\,\mu$m ratio from \textit{Spitzer} and \textit{Herschel} observations of our sample (red), which is drawn from the larger sample of \cite{Kirkpatrick2015}, shown as circles. The parent sample is colored by their \textit{Spitzer}-derived spectroscopic redshifts. Histograms on each axis show the flux ratio distribution for our sample (red) and the parent sample (gray). We are biased toward larger $S_{250}/S_{24}$ ratios because of our redshift selection, but our sample spans 16th and 84th percentiles of the parent distribution along both axes.}
    \label{fig:colorcolor}
\end{figure}

Our final proposed sample consisted of 60 galaxies in GOODS-N/S and xFLS, for which we were awarded time with \textit{JWST} to observe the 40 targets in GOODS-N and xFLS during Cycle 2 (Program \#3224, PI: McKinney). The approved subset is broadly representative of the parent sample with respect to \textit{Spitzer} and \textit{Herschel} colors as shown in Figure \ref{fig:colorcolor}, although we are biased by our redshift selection. Our sample also reproduces the general demographics with respect to the breakdown between purely star-forming galaxies, AGN, and composites. 
The \textit{JWST} data presented in this article were obtained from the Mikulski Archive for Space Telescopes (MAST) at the Space Telescope Science Institute. The specific observations analyzed can be accessed via \dataset[DOI: 10.17909/18v0-8068]{https://doi.org/10.17909/18v0-8068}. 

\begin{figure*}
    \includegraphics[width=\textwidth]{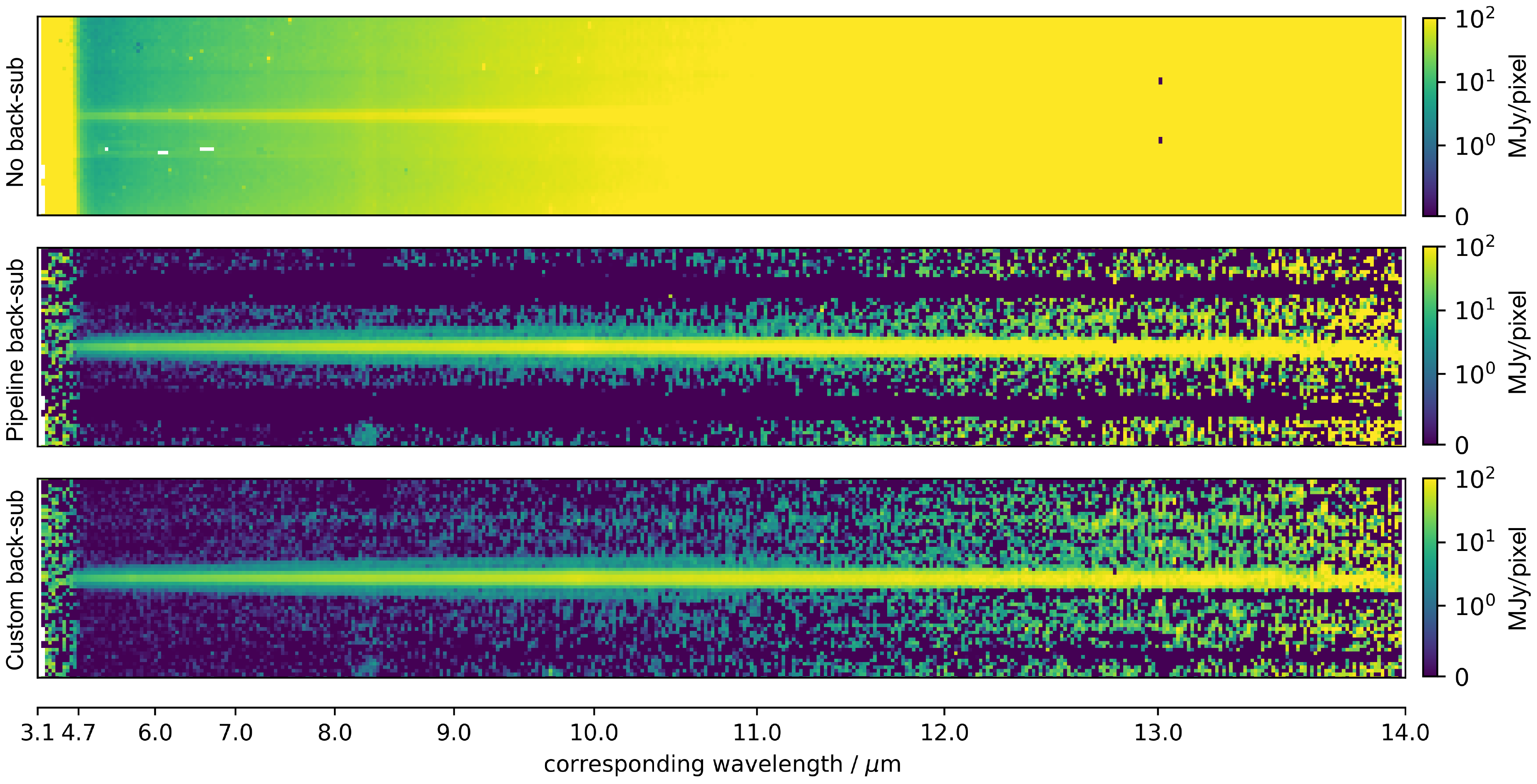}
    \caption{Demonstration of our background subtraction procedure for the 2D spectrum of one of the targets in our sample. The top panel shows the 2D spectrum extracted from the pipeline with no background subtraction, the middle panel shows result of the pipeline's default background subtraction routine, and the bottom panel shows the final product after our two-stage approach as described in Section \ref{sec:data}. In the bottom panel, we show the observed wavelength corresponding to pixels along the spectral axis. We note that the wavelength calibration is not reliable below $4.7\,\mu$m and we do not make use of those data. With this custom background subtraction routine, we are able to extract spectra between $4.7$ and $14\,\mu$m.}
    \label{fig:lrs:bkg}
\end{figure*}

\subsection{Description of Observations\label{sec:obs}}

We observed our targets with the \textit{JWST}/MIRI LRS in the fixed slit mode which covers $\lambda_{obs}=5-14\,\mu$m. LRS slit spectroscopy makes use of a $0.51^{\prime\prime}\times4.7^{\prime\prime}$ slit to achieve $R=\lambda/\Delta\lambda\sim100$ spectra, varying between $R\sim40$ at $5\,\mu$m to R$\sim200$ at $12\,\mu$m \citep{Kendrew2015,Rieke2015}. For all but one of our targets, we performed target acquisition (TA) using the science target and the MIRI imager using F560W or F1500W, depending on which filter the source was expected to be brighter in. We also obtained LRS verification images in F560W, F770W, or F1000W that are used for pointing verification. In the future, the verification images will also be used for extracting LRS wide-field slitless spectra of sources in the main MIRI imager field-of-view. 

The LRS observations were configured in two ways to preserve uniform exposure times across bright and faint targets. We identified ``bright'' ($N=30$) and ``faint'' ($N=10$) targets as having \textit{Spitzer}/IRAC 8$\,\mu$m fluxes above and below 32 $\mu$Jy, respectively. This flux threshold was chosen based on our calculations with the \textit{JWST} Exposure Time Calculator prior to \textit{JWST} Cycle 2. 
Exposure times for targets in the bright subset were based on calculations that assumed the minimum IRAC/8$\,\mu$m$\,=32\,\mu$Jy flux within that subsample. Exposure times for targets in the faint subset were based on calculations that assumed the minimum IRAC/8$\,\mu$m$\,=10\,\mu$Jy flux within that subsample. 


For the bright subset, we performed a single along slit nod (2 dithers) with 81 groups per integration, two integrations per exposure and one exposure per dither. This corresponded to a total exposure time on-source of $904.6$ s. For the faint subset of our sample, we also performed a single along slit nod (two dithers) but with 108 groups per integration, six integrations per exposure, and one exposure per dither. This corresponded to a total exposure time on-source of $3624.2$ s. Every source was observed fully in a single visit. These observations were scheduled and executed between 2024 January and 2024 April. 

Of our 40 targets, five observations failed because on-source target TA failed to centroid using the short ($<300$ s) TA exposures. Three of these were successfully reobserved using a longer TA exposure, and in some cases, a different TA imaging filter. The remaining two TA failures were not approved for reobservations. Finally, one of the targets that had an initially successful observation was not detected because TA centered the LRS slit on a hot pixel off-source. Our final sample consists of 37 galaxies with MIRI LRS spectra. 


\section{\textit{JWST} Data Reduction\label{sec:data}}
\subsection{MIRI/LRS Reduction}
We reduced the MIRI/LRS data using the \textit{JWST} pipeline (version $1.15.1$, \citealt{jwst_pipeline}) with the default settings for Stages 1 and 2 responsible for detector level processing, flat-fielding, photometric calibration, pixel replacement, and spectroscopic resampling. Prior to running Stage 3 of the pipeline, we create empirical backgrounds for each observation by masking out the trace from individual nods and then stacking them. We then carry out a custom background subtraction routine, which we use to subtract the backgrounds in place of the default Stage 3 routines. This tailored approach is detailed below. 

First, we create median backgrounds from observations taken within 2 weeks of one another. Observations separated by more than 2 weeks tend to exhibit significant fluctuations in the overall background levels, typically $\pm30\%$, as well as comparable changes in the prominence and shape of specific artifacts, such as the ``cruciform'' \citep{Wright_2023}, the bar\footnote{The bar artifact in the blue domain is caused by undispersed persistence in the LRS slit from the previous observation. In our case, the previous observations are always the MIRI imaging TA exposure in F560W, F1000W, or F1500W.}, and striping. We find that the use of median backgrounds combining observations taken within 2 weeks of one another yields the best final 2D and 1D spectra. For each individual nod, we median-normalize and then subtract the corresponding ``master'' median background. Then, we execute the Stage 3 pipeline, skipping the default background subtraction routines, to construct the final 2D spectra. We finally implement one last background correction step that accounts for case-by-case variations in the background slope across the slit as well as residual artifact removal. After masking the trace from the 2D spectra, we subtract out the column-averaged residual backgrounds row-by-row, zeroing out the background uniformly across each 2D spectrum. Figure \ref{fig:lrs:bkg} shows one example of our successful background subtraction procedure, compared against the default pipeline approach.

To extract 1D spectra, we implement a PSF-weighted aperture custom-tailored for each source to encase the full extent of the trace. In 50\% of the targets (18/37), the trace is extended for which we manually adjust the aperture to capture the total flux. Sources with extended traces tend to be at lower redshift as expected, with 
$\langle z\rangle=1.35_{-0.55}^{+0.65}$ as compared with $\langle z\rangle =1.95_{-0.75}^{+0.05}$ for the 19 targets with point-like traces. 
We note that these two subsets have similar distributions in \lir, mid-infrared AGN fraction, and final $3.3\,\mu$m PAH line luminosities (as described later in Section \ref{sec:decomp}). 
We sum all of the emission within the apertures and then apply the LRS pixel-to-wavelength calibration\footnote{Updated as of December, 2023 after issues with the original wavelength calibration were identified \citep{Kwok2023}.}. The uncertainties are measured from both the pixel variance along signal-free regions of the 2D spectrum as well as extractions from the corresponding 2D weight maps combined in quadrature. Noise derived from the science maps dominates the error because these mid-infrared observations are background limited. 

\begin{figure}
    \centering
    \includegraphics[width=\linewidth]{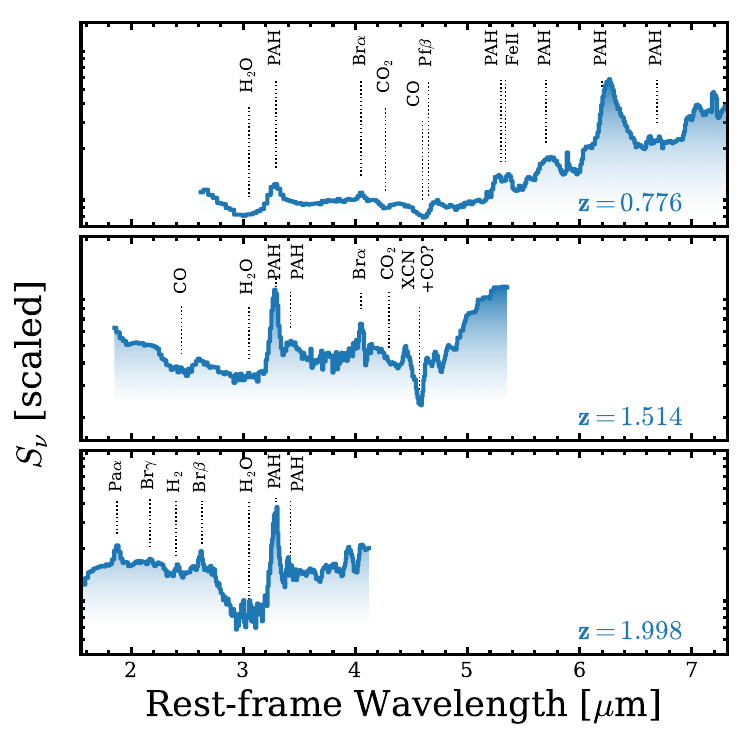}
    \caption{\textit{JWST} MIRI LRS $\lambda_{\rm obs}=5-14\,\mu$m spectra of $z\sim1-2$ mid-infrared--bright galaxies. We highlight three star-forming galaxies at $z_{\rm spec}=0.776$ (\textit{Top}), $z_{\rm spec}=1.515$ (\textit{Middle}), and $z_{\rm spec}=1.998$ (\textit{Bottom}). Prominent $3.05\,\mu$m H$_2$O ice absorption can be seen in all three spectra, including other PAH features as well as atomic lines such as Pa$\alpha$ and Br$\alpha$. 
    }
    \label{fig:lrs:sfgs}
\end{figure}

\begin{figure}
    \centering
    \includegraphics[width=\linewidth]{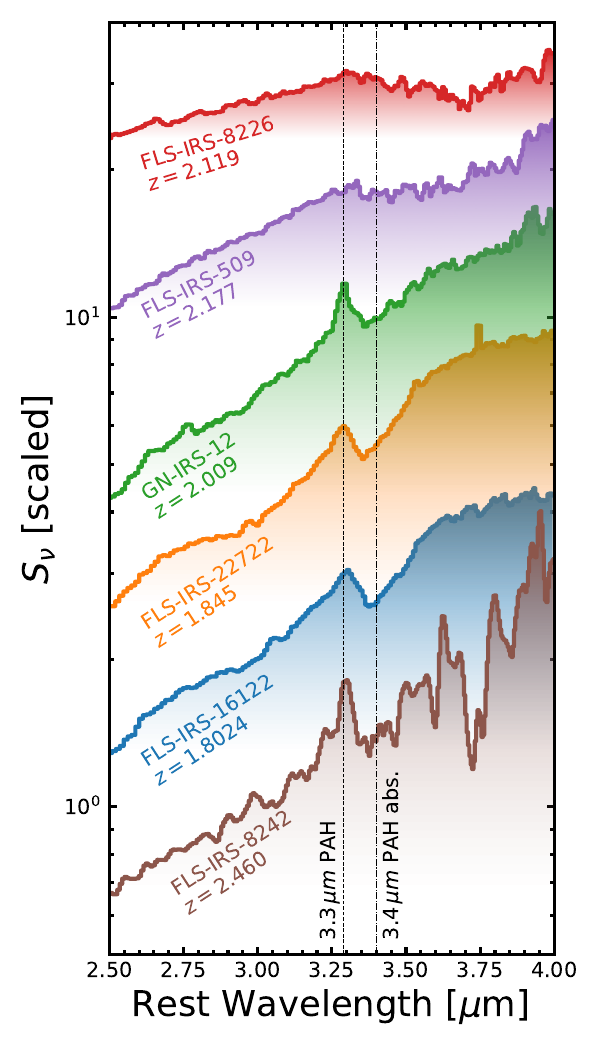}
    \caption{\textit{JWST} MIRI LRS spectra of six mid-infrared--bright AGN at $z=1.8-2.4$. 
    These AGN spectra are dominated by a rising continuum originating from hot dust. The $3.3\,\mu$m PAH, marked with a vertical dashed line, is detected in 4/6 of these power-law AGN. Aliphatic absorption is seen at $3.4\,\mu$m (dotted--dashed line) in FLS-IRS-22722, FLS-IRS-16122, and GN-IRS-12.
    }
    \label{fig:lrs:agns}
\end{figure}

The final 1D spectra achieve a median SNR per spectral element $\approx50$ at $\lambda_{\rm obs}=5\,\mu$m, SNR\,$\approx25$ at $\lambda_{\rm obs}=9\,\mu$m, and SNR\,$\approx5$ at $\lambda_{\rm obs}=13\,\mu$m. Figure \ref{fig:lrs:sfgs} highlights the various atomic and molecular emission lines we detect in star-forming galaxy targets, as well as the PAHs and ice absorption features. Figure \ref{fig:lrs:agns} highlights the rising spectra observed in the six AGN in our sample with steep rest-frame near-infrared slopes $S_{4.8\,\mu m}/S_{2.8\,\mu m}>1.5$ and the largest contribution of hot dust to their LRS spectrum, where we see PAHs in emission and dust in absorption. 

\subsection{MIRI/LRS Slit-loss Corrections}
The archival \textit{Spitzer} data on our sources includes IRAC $3.6-8\,\mu$m imaging, as well as (sometimes overlapping) \textit{Spitzer} IRS spectra and $16\,\mu$m Peak Up imaging. As discussed in \cite{Kirkpatrick2015}, the IRS spectra have already been corrected for slit losses based on \textit{Spitzer} photometry, but these are negligible owing to the large 10$^{\prime\prime}$ width of the IRS slit. 
We use these observations to correct the MIRI/LRS spectra assuming the emission caught in the $0.51^{\prime\prime}\times4.7^{\prime\prime}$ slit is representative of the galaxy-integrated spectrum. 

For galaxies in the mass and $A_V$ range of our sample, $\log M_*/M_\odot>10.5$ and $A_V>1$ \citep{Choi2006,Rujopakarn2012,Kirkpatrick2012}, \textit{JWST} MIRI imaging finds mid-infrared effective radii to be $\approx2$ kpc \citep{Magnelli2023,Lyu2025primer}. At $z\sim1-2$, this corresponds to a total extent of $\approx0.5^{\prime\prime}$ which is already at the LRS slit width without accounting for PSF effects. Knowing that the FWHM of the PSF over LRS' wavelength ranges from $0.2^{\prime\prime}-0.4^{\prime\prime}$, we can expect slit-loss corrections to be important. 
We calculate these correction factors using the ratio of \textit{Spitzer} IRAC 5.6$\,\mu$m and 8$\,\mu$m photometry to the observed LRS spectrum convolved with the corresponding filter transmission profile. We use the IRS spectra and $16\,\mu$m imaging for consistency checks and find that corrected LRS spectra nicely align within the uncertainties of both. The median slit-loss correction factor is $4$ with upper and lower quartiles of $2$ and $8$, respectively. 

\begin{figure*}
\centering
\includegraphics[width=0.3\textwidth]{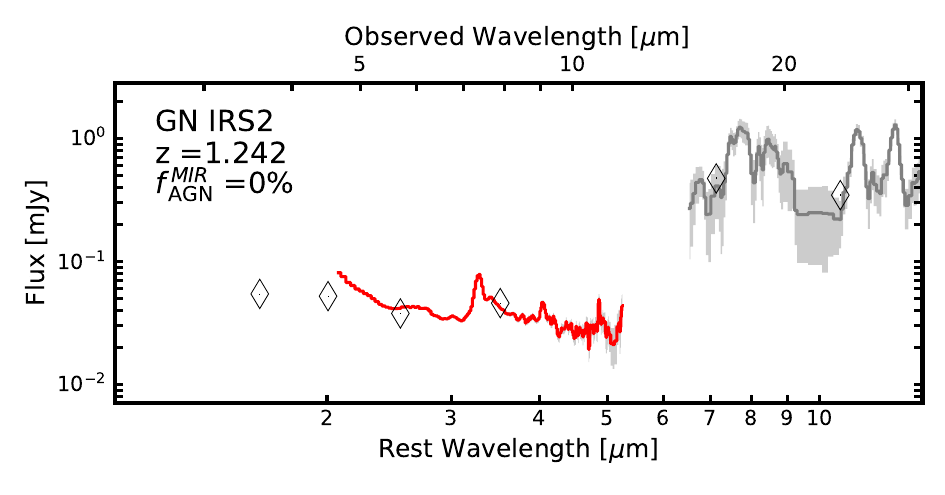}
\includegraphics[width=0.3\textwidth]{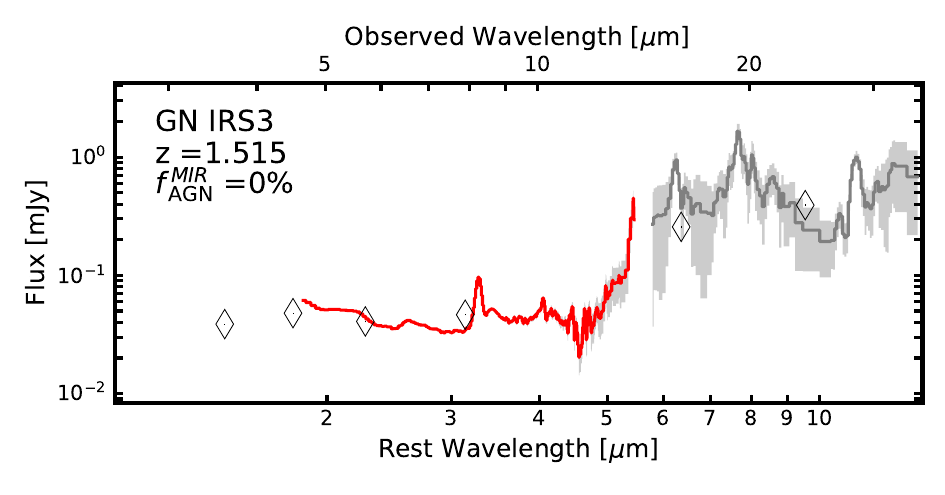}
\includegraphics[width=0.3\textwidth]{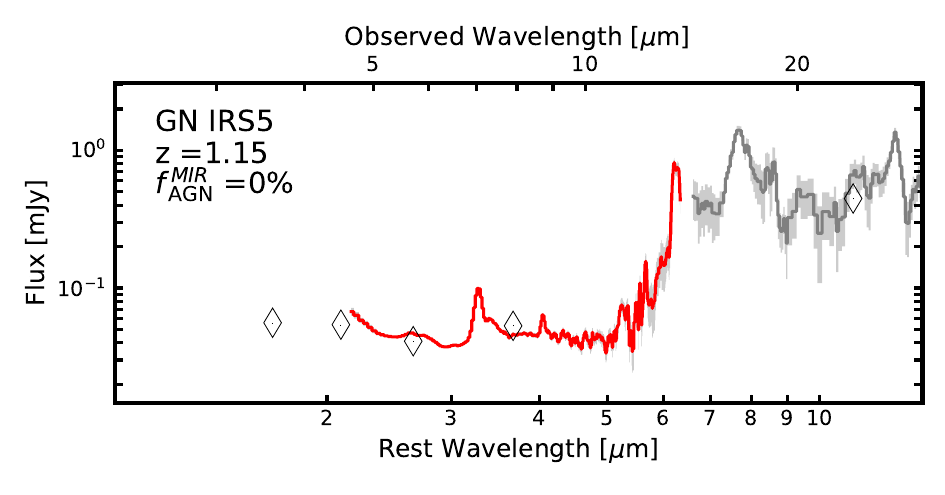}
\includegraphics[width=0.3\textwidth]{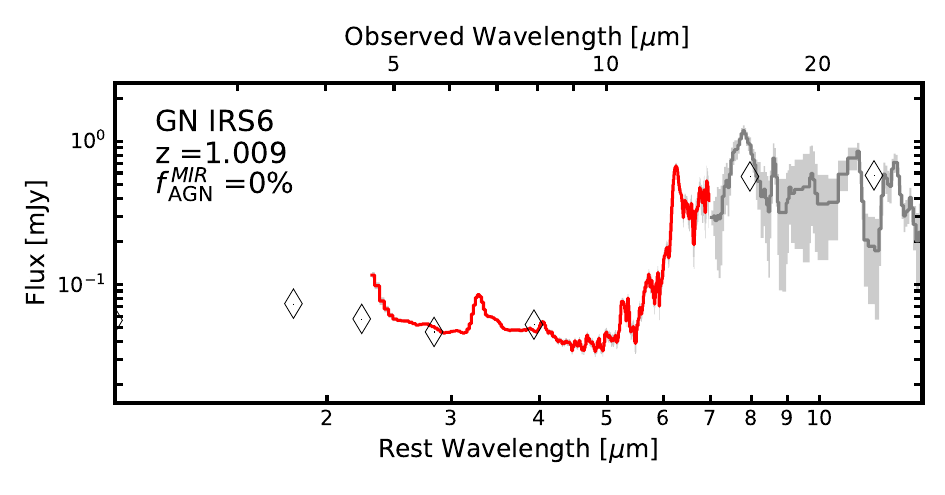}
\includegraphics[width=0.3\textwidth]{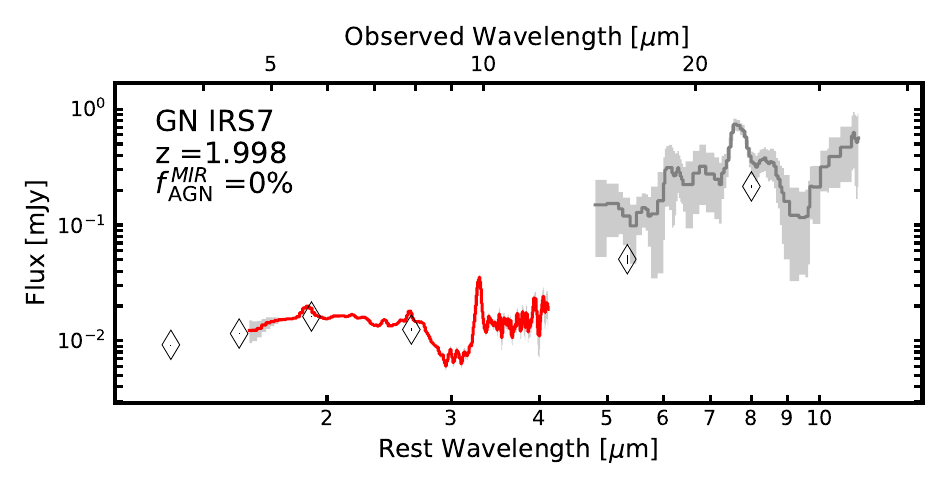}
\includegraphics[width=0.3\textwidth]{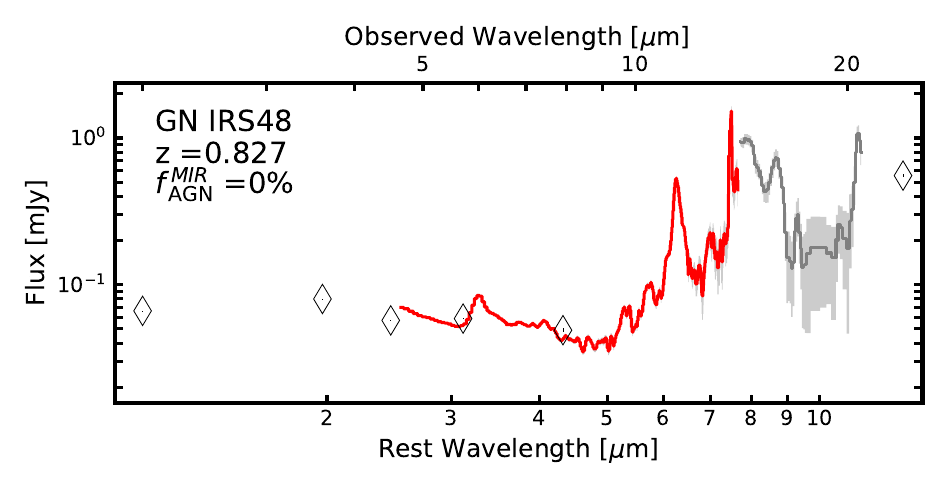}
\includegraphics[width=0.3\textwidth]{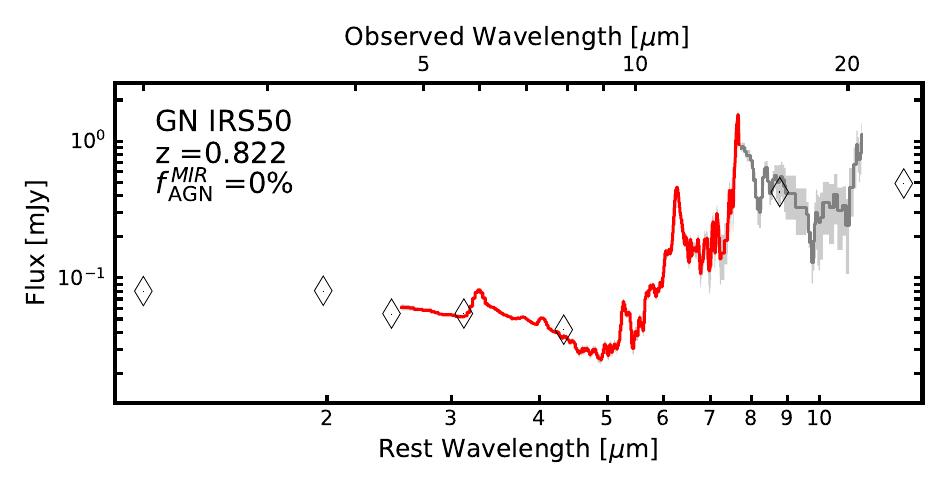}
\includegraphics[width=0.3\textwidth]{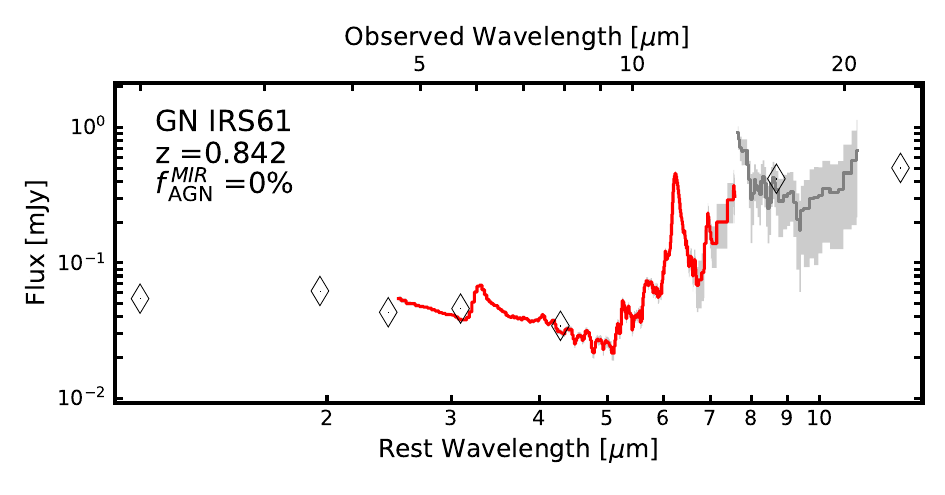}
\includegraphics[width=0.3\textwidth]{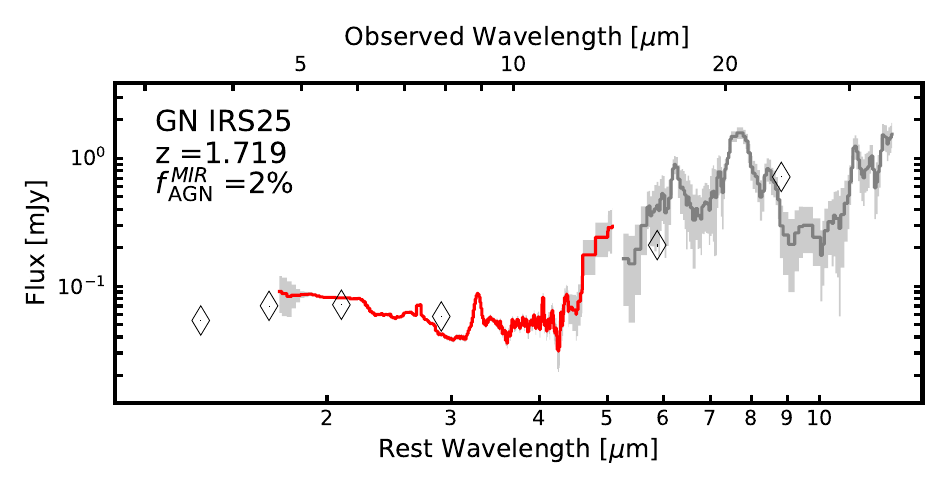}
\includegraphics[width=0.3\textwidth]{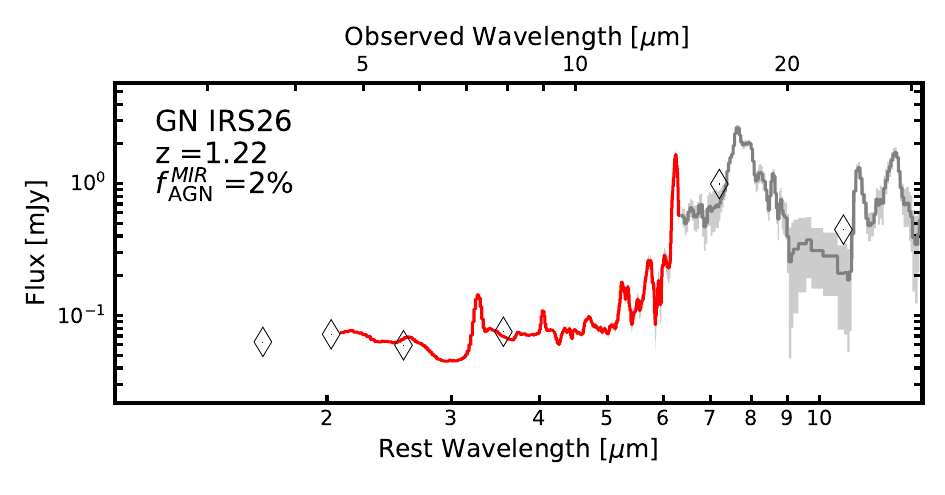}
\includegraphics[width=0.3\textwidth]{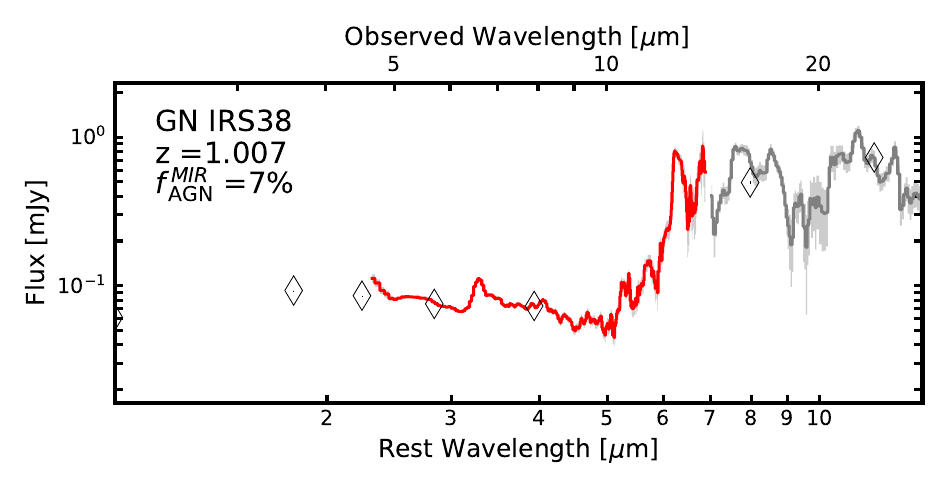}
\includegraphics[width=0.3\textwidth]{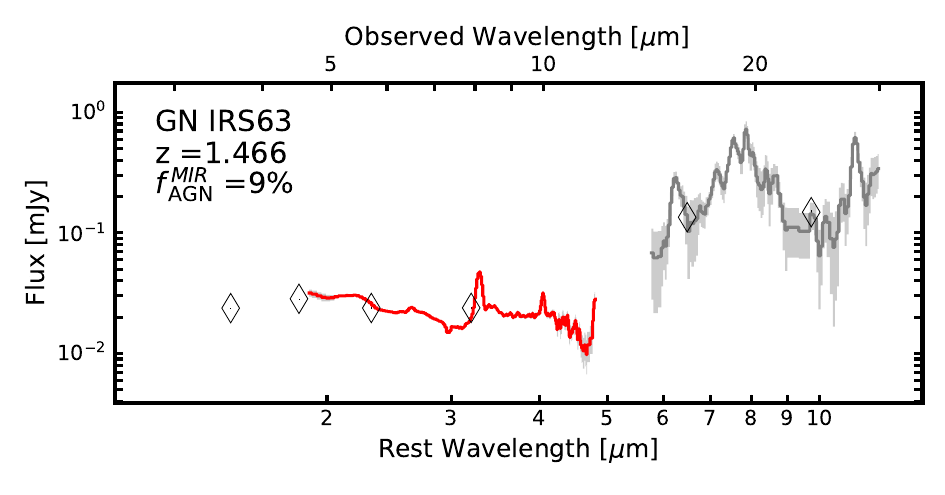}
\caption{Combined \textit{JWST}/MIRI LRS (red) and \textit{Spitzer}/IRS (gray) spectra spanning $\lambda_{\rm rest}\sim2-15\,\mu$m ($\lambda_{\rm obs}\sim5-30\,\mu$m) for the galaxies in our sample with $f_{\rm AGN}<10\%$. Black diamonds show the \textit{Spitzer} photometry between $5.6$ and $24\,\mu$m from IRAC, MIPS, and IRS Peak Up imaging (where available). We detect the $3.3\,\mu$m PAH feature between $z=0.8-2$ in massive, star-forming galaxies with longer wavelength PAHs from the \textit{Spitzer} spectra. The $3.05\,\mu$m ice absorption feature is always present blueward of the $3.3\,\mu$m PAH. }
\label{fig:specs:sfg}
\end{figure*}

\begin{figure*}
\centering
\includegraphics[width=0.3\textwidth]{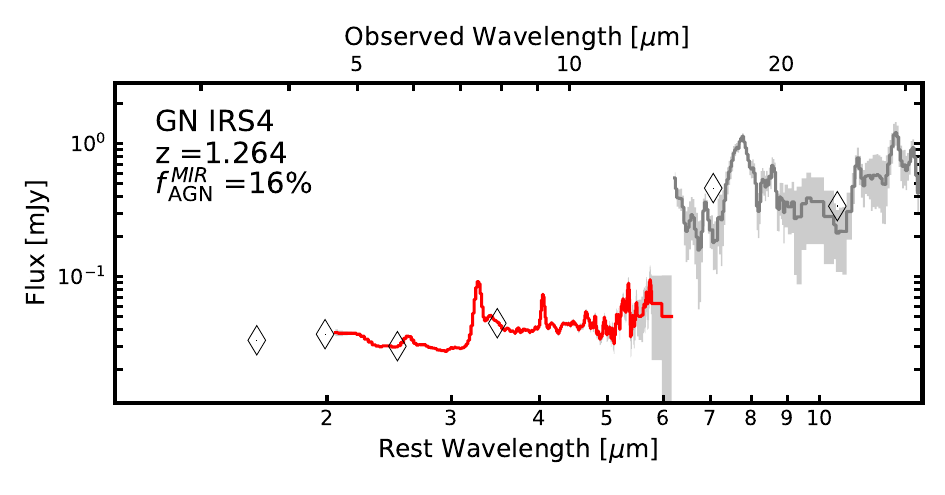}
\includegraphics[width=0.3\textwidth]{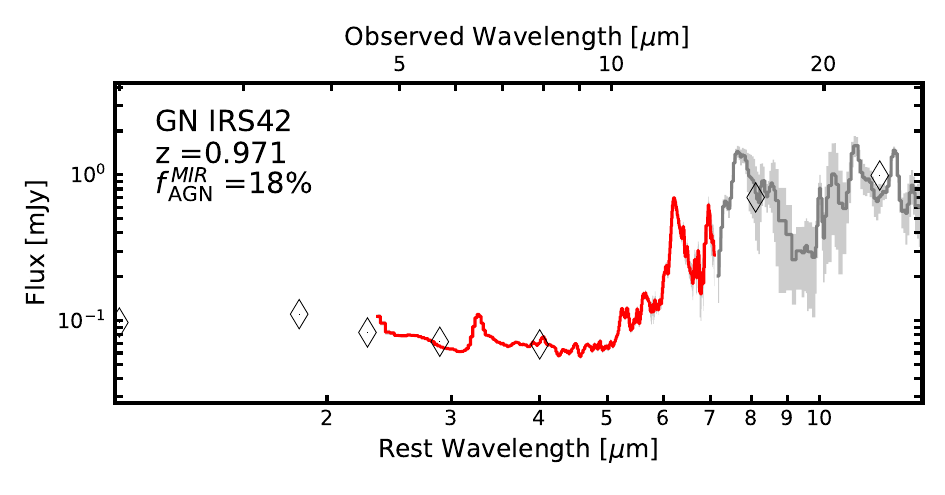}
\includegraphics[width=0.3\textwidth]{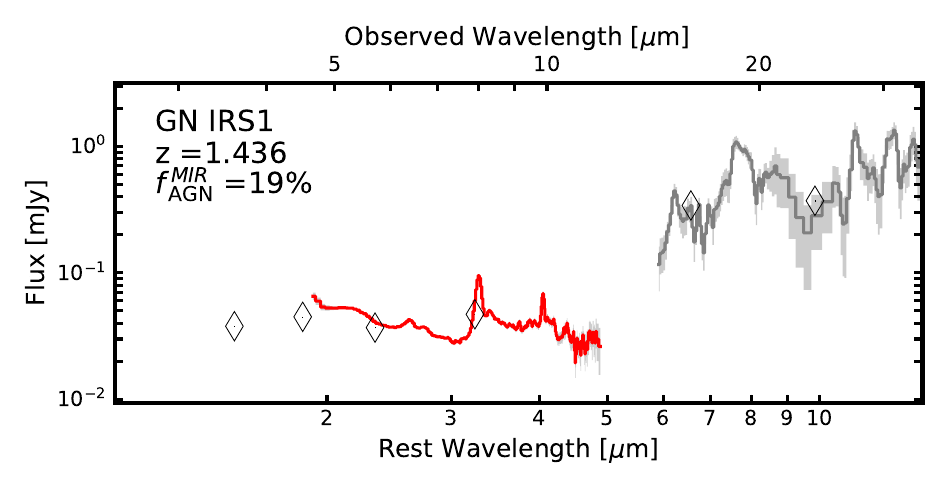}
\includegraphics[width=0.3\textwidth]{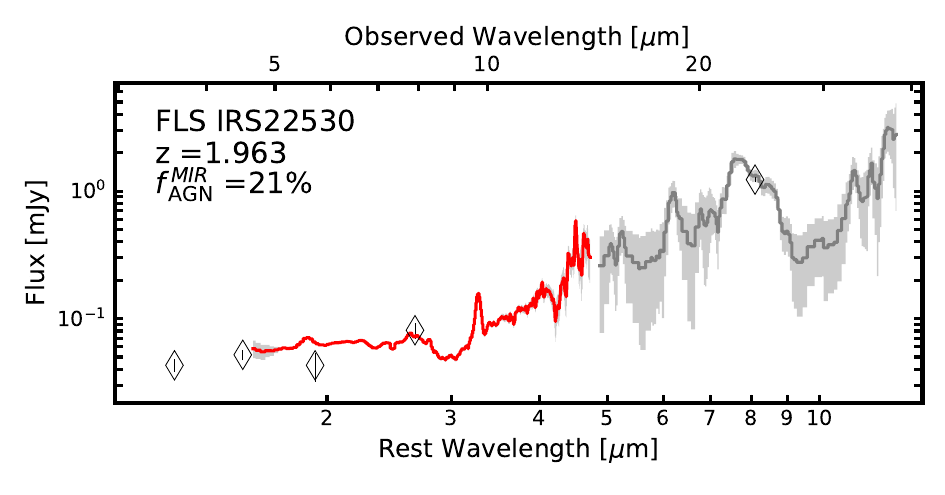}
\includegraphics[width=0.3\textwidth]{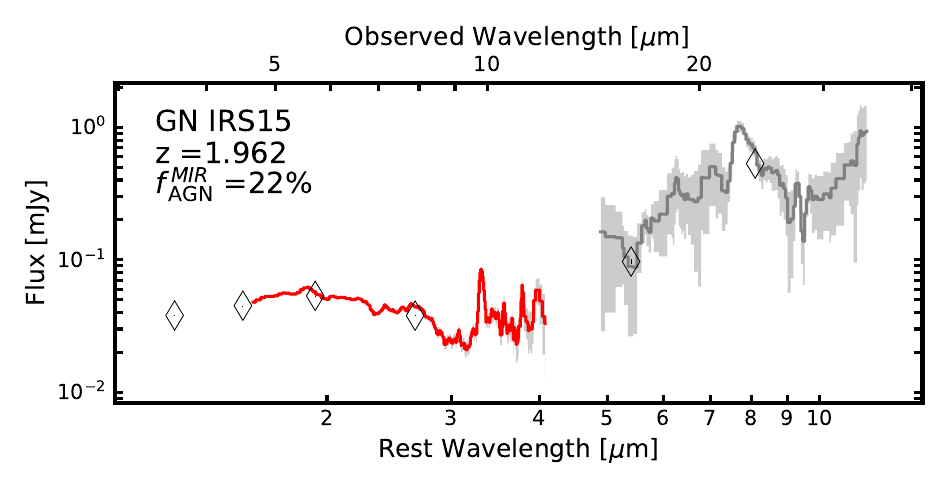}
\includegraphics[width=0.3\textwidth]{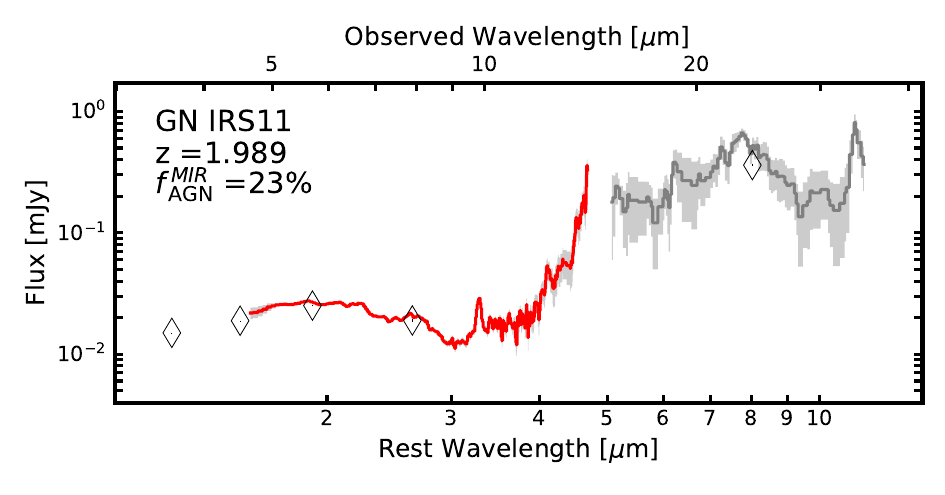}
\includegraphics[width=0.3\textwidth]{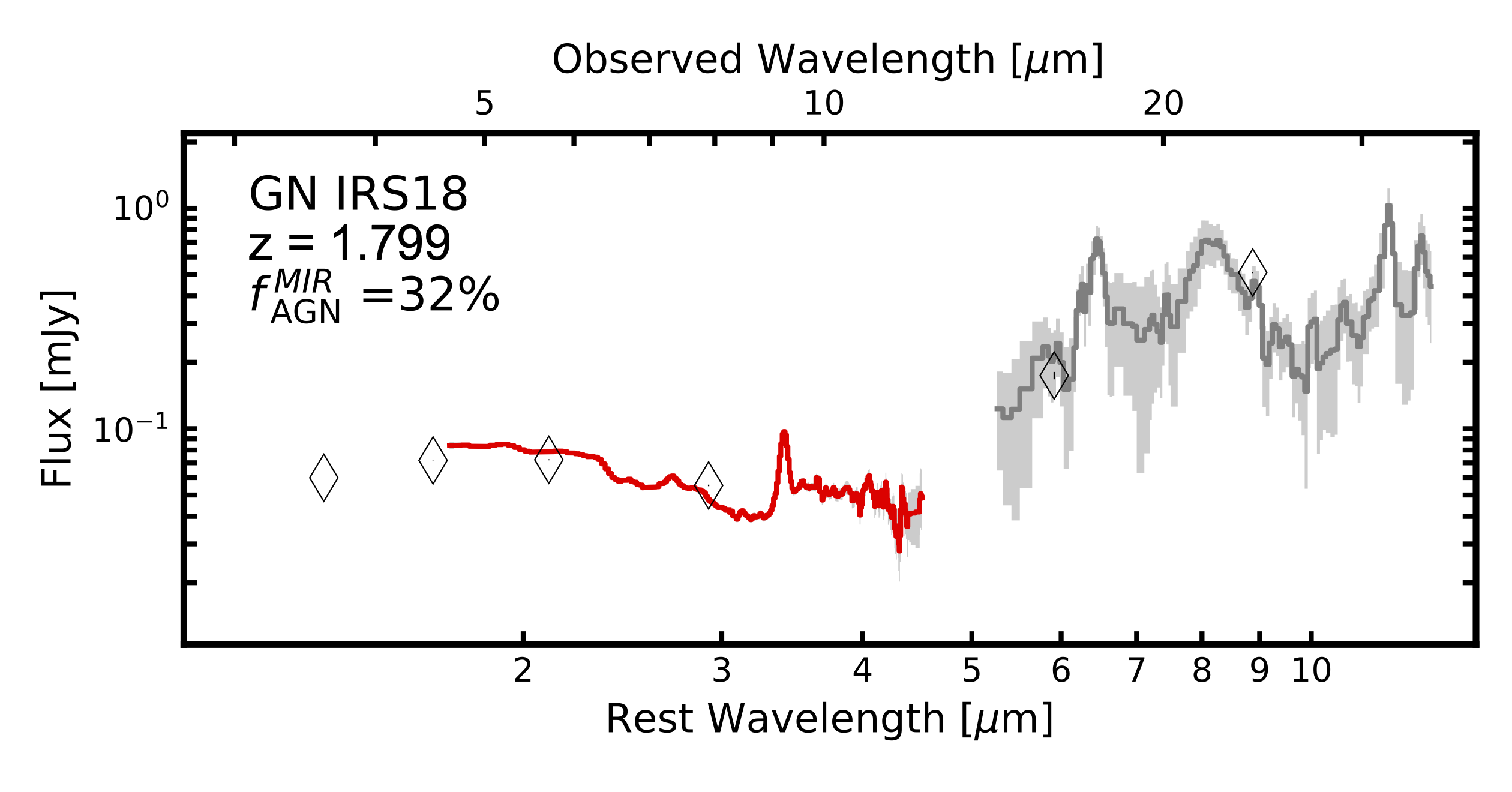}
\includegraphics[width=0.3\textwidth]{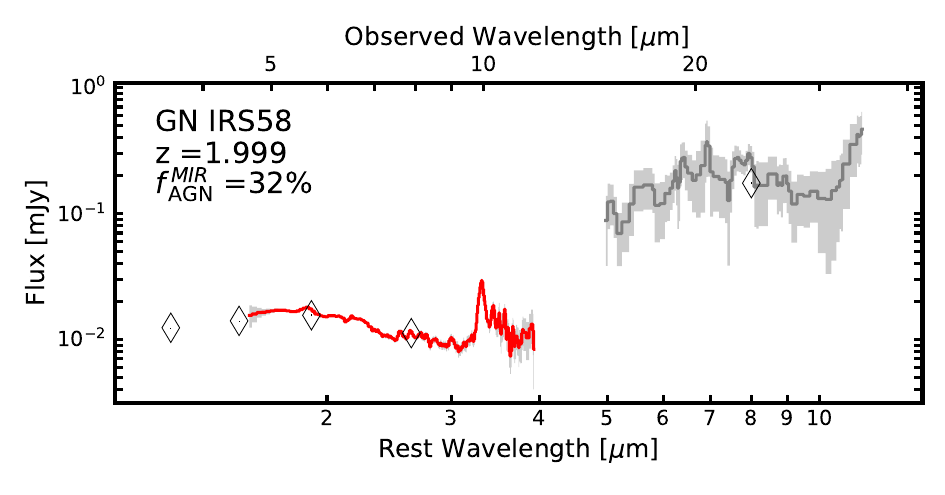}
\includegraphics[width=0.3\textwidth]{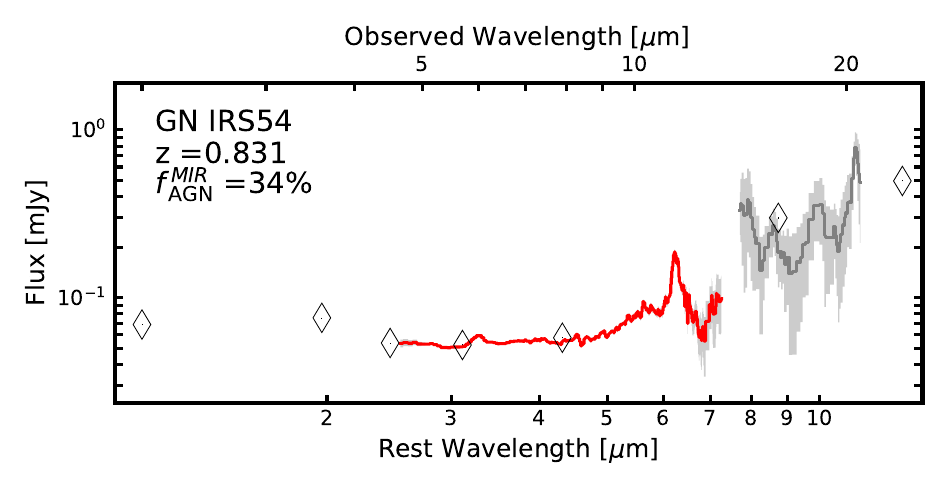}
\includegraphics[width=0.3\textwidth]{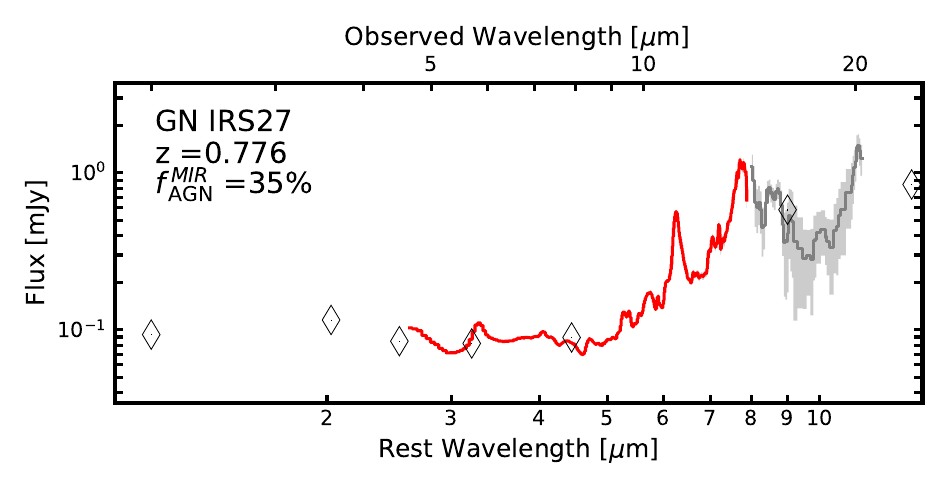}
\includegraphics[width=0.3\textwidth]{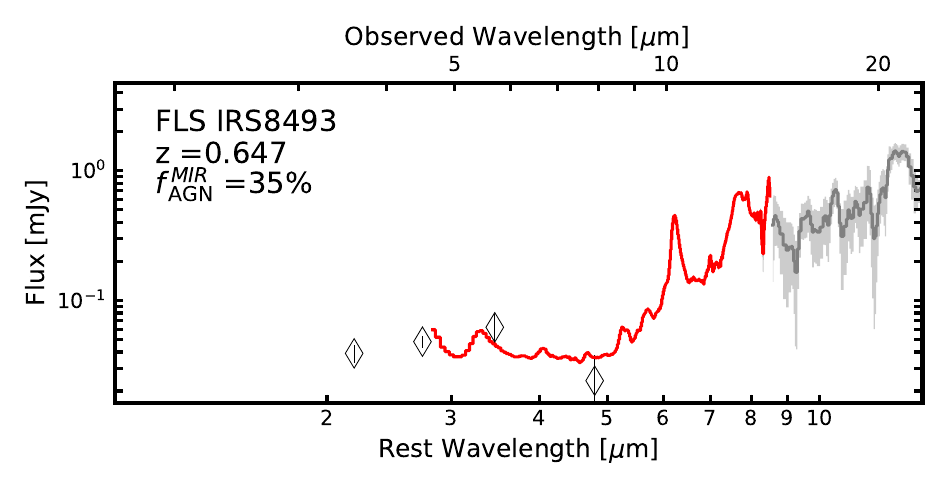}
\includegraphics[width=0.3\textwidth]{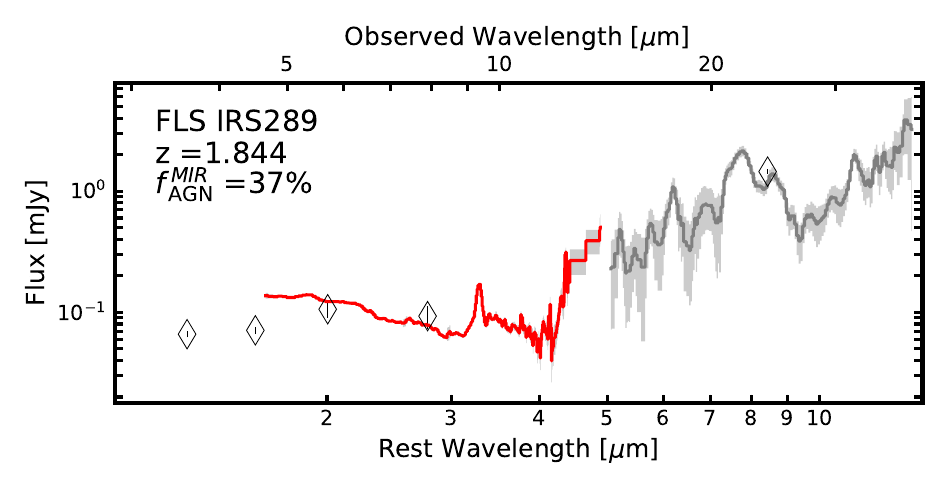}
\includegraphics[width=0.3\textwidth]{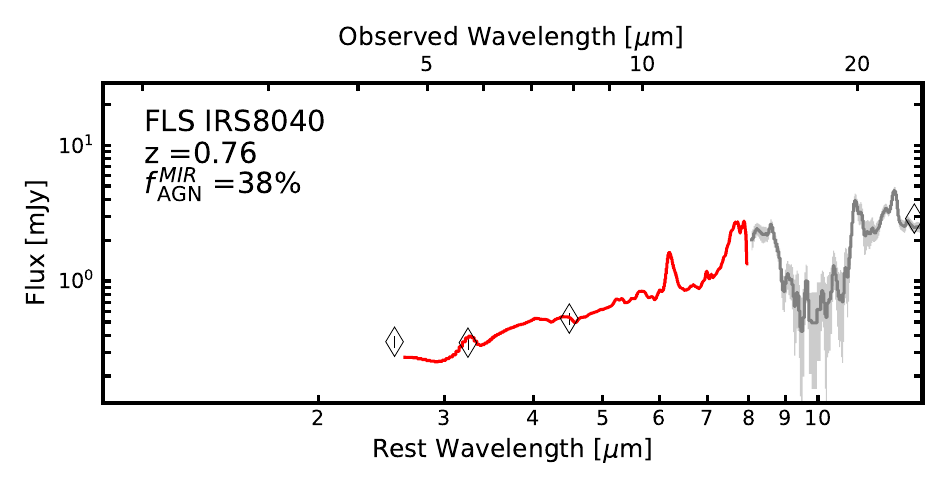}
\includegraphics[width=0.3\textwidth]{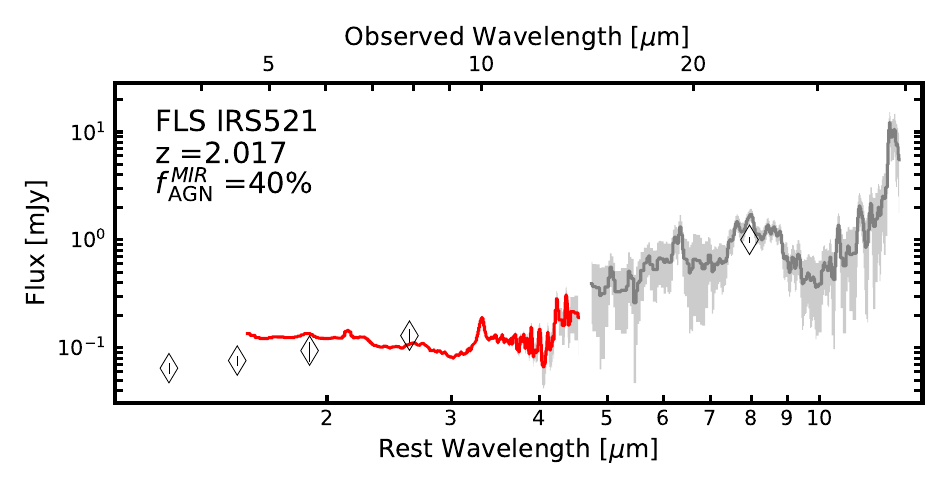}
\caption{Same as Figure \ref{fig:specs:sfg} but for galaxies in our sample with $10\%<f_{\rm AGN}<40\%$. Strong PAH features are still present across this subset of our sample, but the rising continuum from hot dust between $\lambda_{\rm rest}=3-6\,\mu$m becomes more apparent. }
\label{fig:specs:comp}
\end{figure*}

\begin{figure*}
\centering
\includegraphics[width=0.3\textwidth]{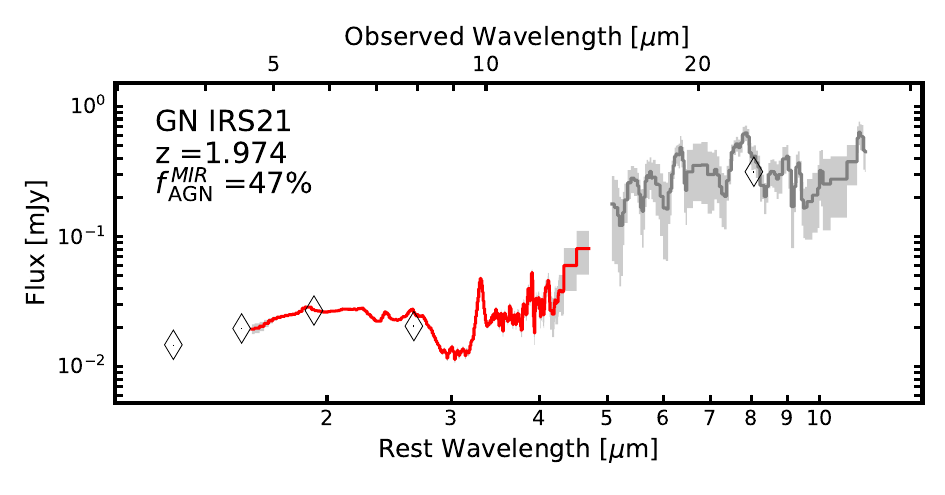}
\includegraphics[width=0.3\textwidth]{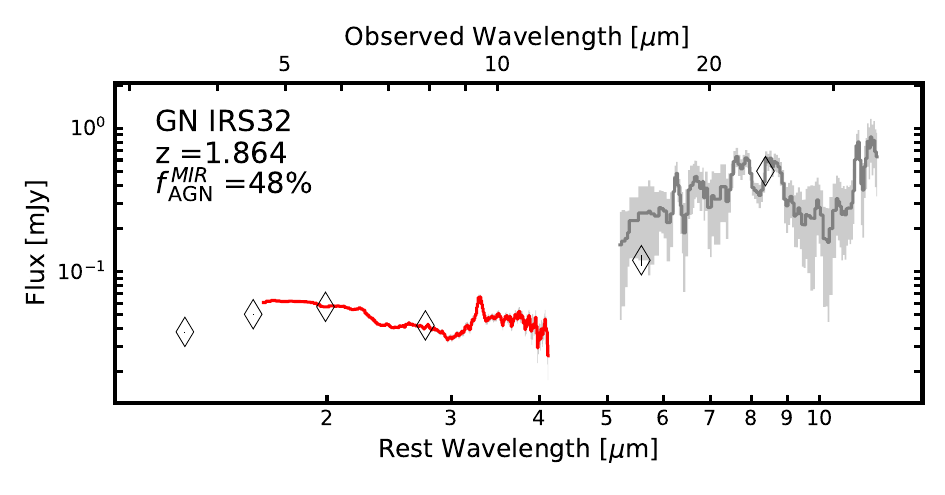}
\includegraphics[width=0.3\textwidth]{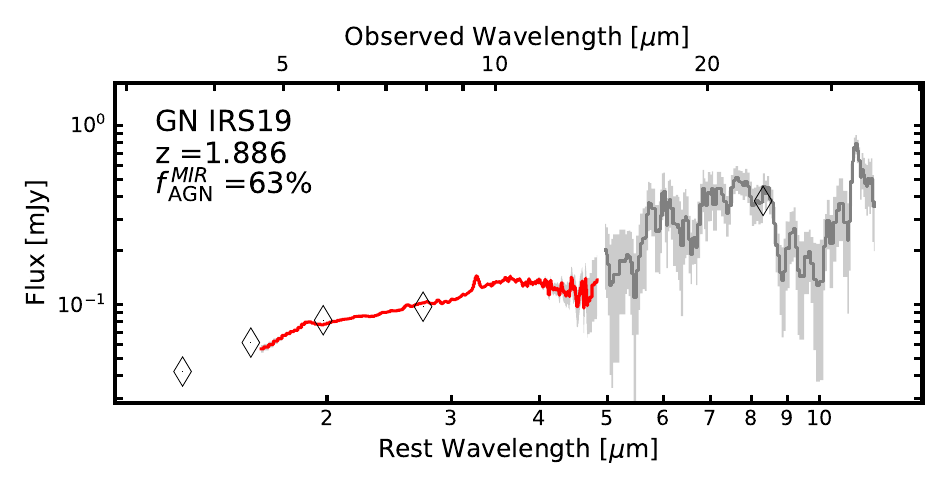}
\includegraphics[width=0.3\textwidth]{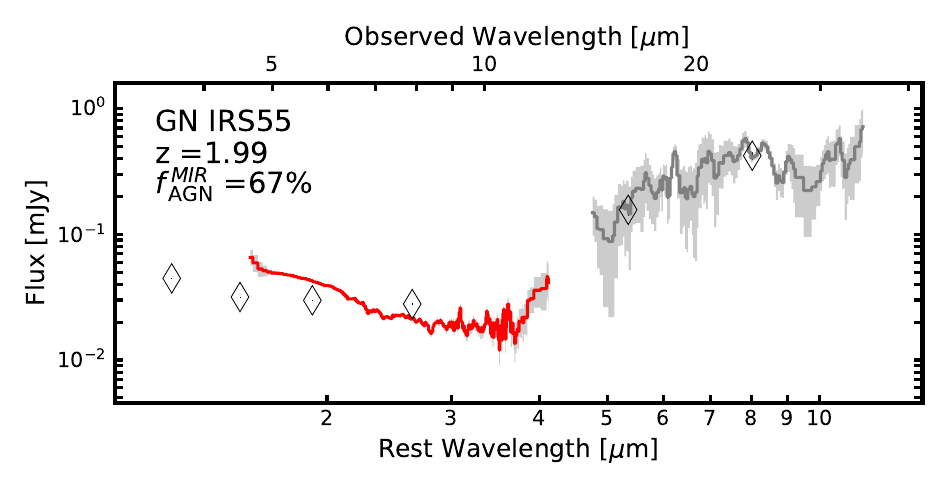}
\includegraphics[width=0.3\textwidth]{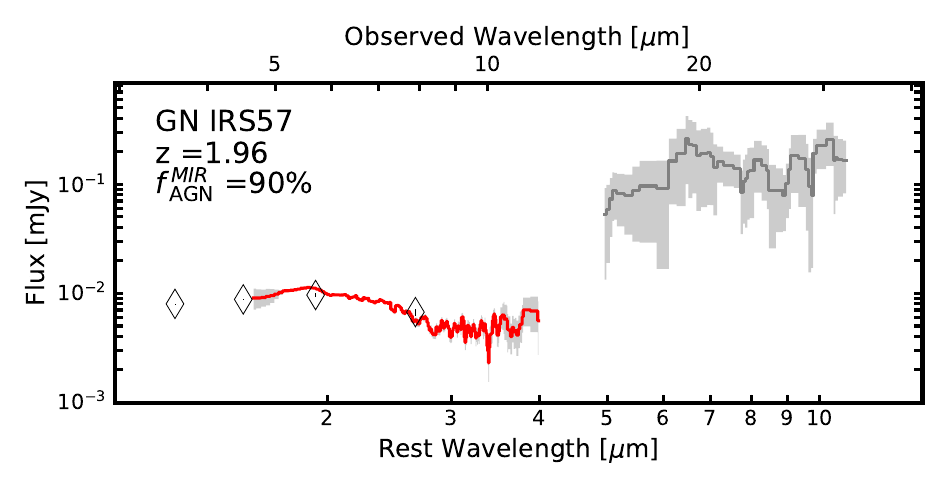}
\includegraphics[width=0.3\textwidth]{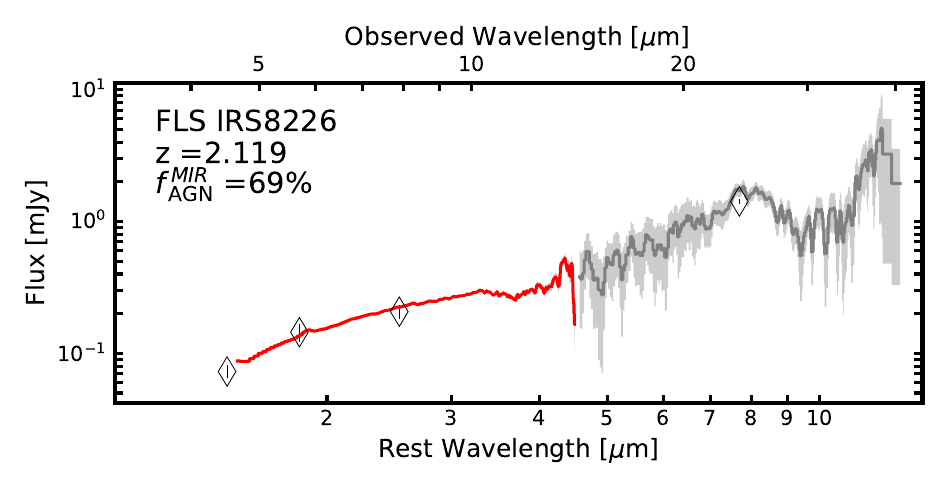}
\includegraphics[width=0.3\textwidth]{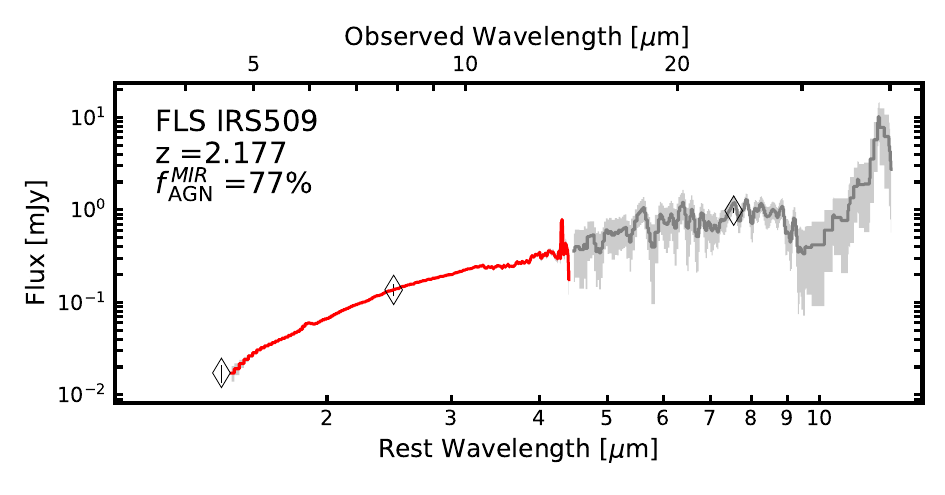}
\includegraphics[width=0.3\textwidth]{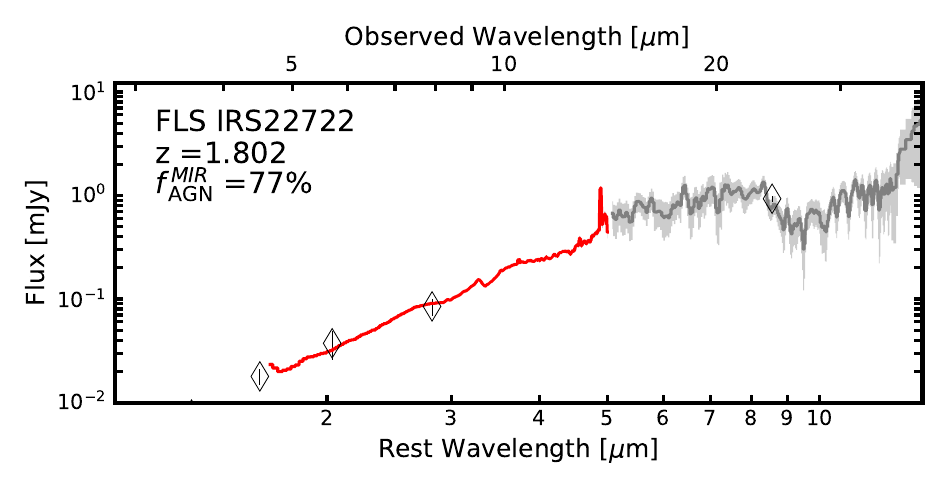}
\includegraphics[width=0.3\textwidth]{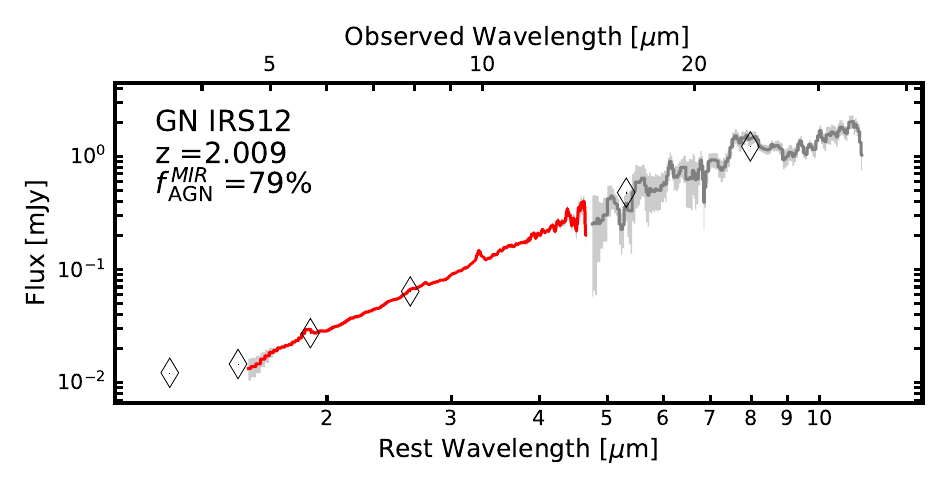}
\includegraphics[width=0.3\textwidth]{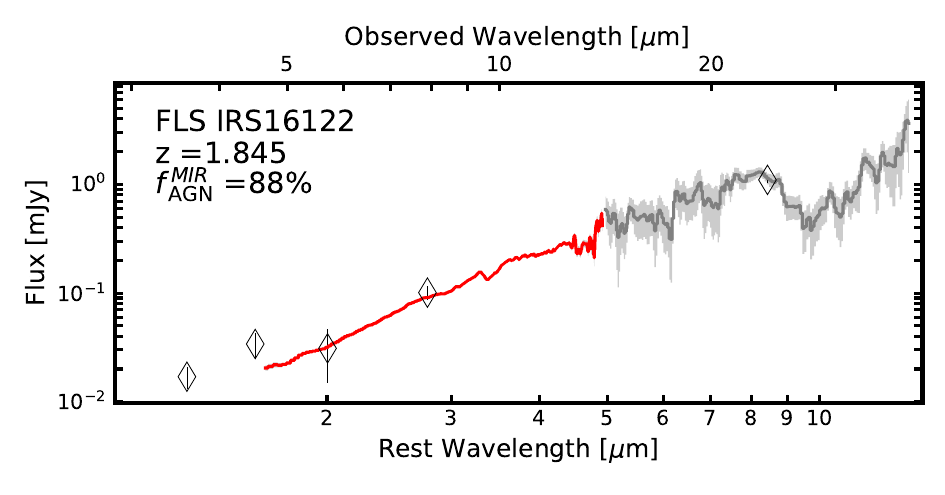}
\includegraphics[width=0.3\textwidth]{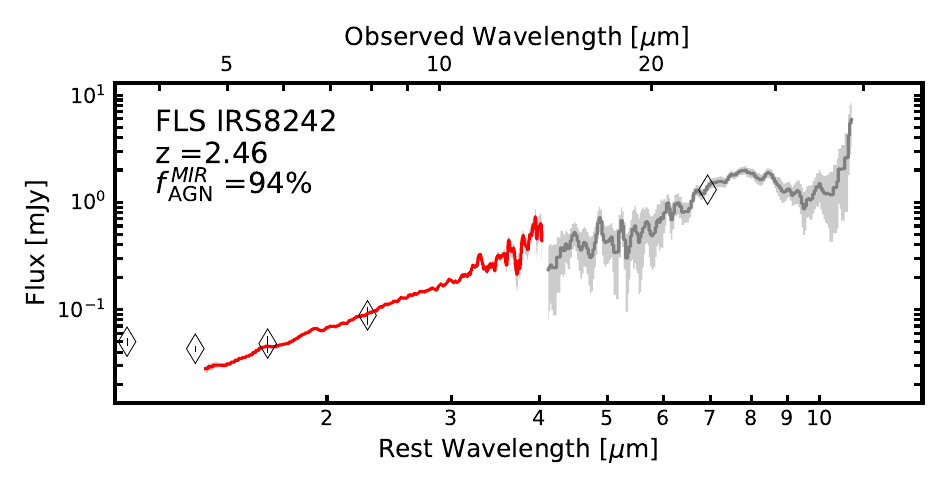}
\caption{Same as Figure \ref{fig:specs:sfg} but for galaxies in our sample with $f_{\rm AGN}>40\%$. Steep hot dust continuum begins to dominate the spectrum from $\lambda_{\rm rest}=2-20\,\mu$m and the strengths of PAH features diminish relative to continuum. }
\label{fig:specs:agn}
\end{figure*}


\section{Analysis\label{sec:analysis}}

Following the LRS slit-loss corrections outlined in \ref{sec:data}, the aligned \textit{JWST}/MIRI LRS and \textit{Spitzer}/IRS spectra achieve $\lambda_{\rm obs}\sim5-30\,\mu $m spectral coverage for 37 galaxies. We present these combined spectra in Figure \ref{fig:specs:sfg} for star-forming galaxies with mid-infrared AGN fractions ($f_{\rm AGN}$) $f_{\rm AGN}<10\%$ \citep{Kirkpatrick2015}. In brief, $f_{\rm AGN}$ come from \cite{Kirkpatrick2015} and are measured by fitting the \textit{Spitzer} spectra with a power-law continuum representing hot AGN torus dust and a pure star-forming galaxy template \citep[e.g.,][]{Pope2008}. Then, $f_{\rm AGN}$ is defined as the ratio of the integrated power-law component to the total mid-infrared flux between $\lambda_{\rm rest}\sim5-18\,\mu$m. 
Figure \ref{fig:specs:comp} shows the  combined spectra for galaxies with $10\%<f_{\rm AGN}<40\%$, and Figure \ref{fig:specs:agn} shows the mid-infrared AGN-dominated spectra. In this section, we discuss fits to these spectra to extract atomic lines, PAH feature fluxes, and equivalent widths (EWs), as well as the opacities for various ice absorption features. 

\subsection{Spectroscopic Redshifts}
We measure spectroscopic redshifts from the \textit{JWST}/MIRI LRS spectra. In most cases (26/37) we infer $z_{\rm spec}$ from Br$\alpha$ or Pa$\alpha$ and in one instance Pf$\gamma$ which are collectively accurate to $\Delta z\sim10^{-3}$. Nine of the MIRI LRS spectra do not show detectable atomic lines but exhibit bright, high signal-to-noise detections of the 3.3$\,\mu$m PAH and sometimes the $3.4\,\mu$m aliphatic complex. These nine spectra span the redshift range $z=0.8-1.9$ and the full range of $f_{\rm AGN}$, and we measure their spectroscopic redshifts from the PAH features which are still accurate to $\Delta z\sim2\times10^{-3}$. Two AGN-dominated sources show no PAHs, no atomic lines, but absorption features are present (GN-IRS-55 and GN-IRS-57). For these evidently highly obscured sources, we infer the redshifts from the absorption lines. For the majority (36/37) of the sample LRS-derived spectroscopic redshifts agree with the \textit{Spitzer}/IRS redshifts within 5\%. One target (FLS-IRS-8493) at $z_{\rm spec,LRS}=0.647$ was thought to have $z=1.8$ from \textit{Spitzer}, which was based on a misidentification of the $17\,\mu$m PAH feature as the $11.2\,\mu$m complex. LRS-derived redshifts are listed in Table \ref{tab:props}. 


\subsection{Mid-infrared Decomposition\label{sec:decomp}}
The combination of numerous continuum, emission lines, and absorption features at mid-infrared wavelengths makes the details of extracting line fluxes nuanced and important \citep{Smith2007}. 

The ``state-of-the-art'' for mid-infrared spectroscopic decomposition has been developed and optimized for the analysis of low-redshift galaxies with high SNR spectra but similar spectral resolution to LRS ($R\sim100$). Software packages such as \texttt{PAHFIT} \citep{Smith2007,Lai2020} and \texttt{CAFE} \citep{Marshall2007,cafe} leverage the high-quality data available for bright, nearby targets to constrain models with high degrees of freedom. These models include prescriptions for the various dust and atomic emission line features, silicate and ice opacities, and continuum from stars, as well as cool, warm, and hot dust. As a result, there are hundreds of free model parameters that are suitable for such low-redshift observations, especially for data from the \textit{JWST}/MIRI MRS \citep[e.g.,][]{U2022,Lai2022,Lai2023}. Applied to higher-redshift observations with $R\sim100$ where the spectral SNR and wavelength coverage declines, simplifications are required. 

Prior to the launch of \textit{JWST}, the detailed decomposition of mid-infrared spectra for high-redshift galaxies was limited to the brightest and broadest PAH lines in infrared-luminous targets such as submillimeter galaxies and obscured AGN \citep{Houck2005,Weedman2006,Sajina2007,Sajina2009,Yan2005,Yan2007,MenendezDelmestre2007,Pope2008}. The luminosities of the PAH lines were commonly measured using an estimate of the local continuum, which is known to underestimate the intrinsic luminosity of PAHs because of (a) their broad profiles \citep{Smith2007} and (b) extinction effects anchored by the high $9.7\,\mu$m silicate opacities found at $z\sim1-2$ \citep[e.g.,][]{Kirkpatrick2015,Stierwalt2013,Stierwalt2014}. In this work, we infer PAH luminosities in a manner that compromises between the simplicity needed for incorporating lower SNR \textit{Spitzer} IRS data and the flexibility unlocked by the exquisite SNR of the new \textit{JWST} MIRI LRS spectra. Our model is optimized for the joint fitting of LRS and IRS spectra in dust-obscured star-forming galaxies and AGN at $z\sim1-2$ to extract PAH luminosities and opacities.

We model the mid-infrared spectra of our targets following the general outline of \texttt{CAFE} \citep{Marshall2007,Lai2020,cafe} 
as a combination of stellar continuum, thermal dust continuum, and PAH emission features that are corrected for extinction assuming a uniformly mixed medium. The intensity is therefore
\begin{equation}
    I_\nu = \Big[I_\nu^*+I_\nu^{\rm lines}+I_\nu^{\rm dust}+I_\nu^{\rm PAH} \Big]\frac{(1-e^{-\tau_\lambda})}{\tau_\lambda}
\end{equation}
where $I_\nu^*$ and $I_\nu^{\rm dust}$ are continuum from stars and thermal dust respectively, $I_\nu^{\rm lines}$ is atomic and molecular line emission, and $I_\nu^{\rm PAH}$ is the sum of the individual PAH line intensities modeling their profiles as a Lorentzian \citep{Draine2001}. $I_\nu^{\rm lines}$ are only included in overlap regions with the LRS data for sensitivity reasons. The opacity law $\tau_\lambda$ is calibrated on dust in the Milky Way and consists of a power-law component and the $9.7\,\mu$m silicate feature, following \cite{Smith2007}. As in \cite{Lai2020}, we also add the $3.05\,\mu$m ice feature with an optical depth that varies independently. 
The model is convolved with the wavelength-dependent line spread function for MIRI LRS and \textit{Spitzer} IRS. 

\begin{figure*}
    \includegraphics[width=0.49\textwidth]{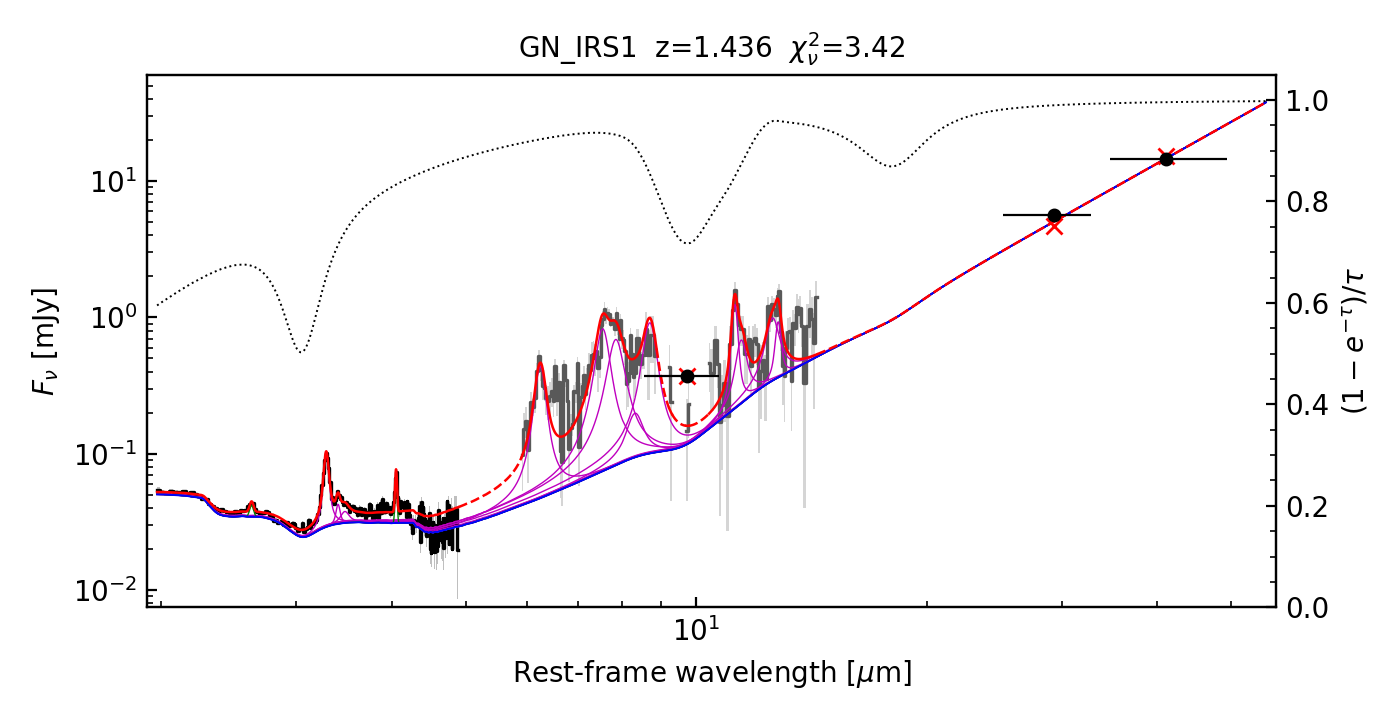}
    \includegraphics[width=0.49\textwidth]{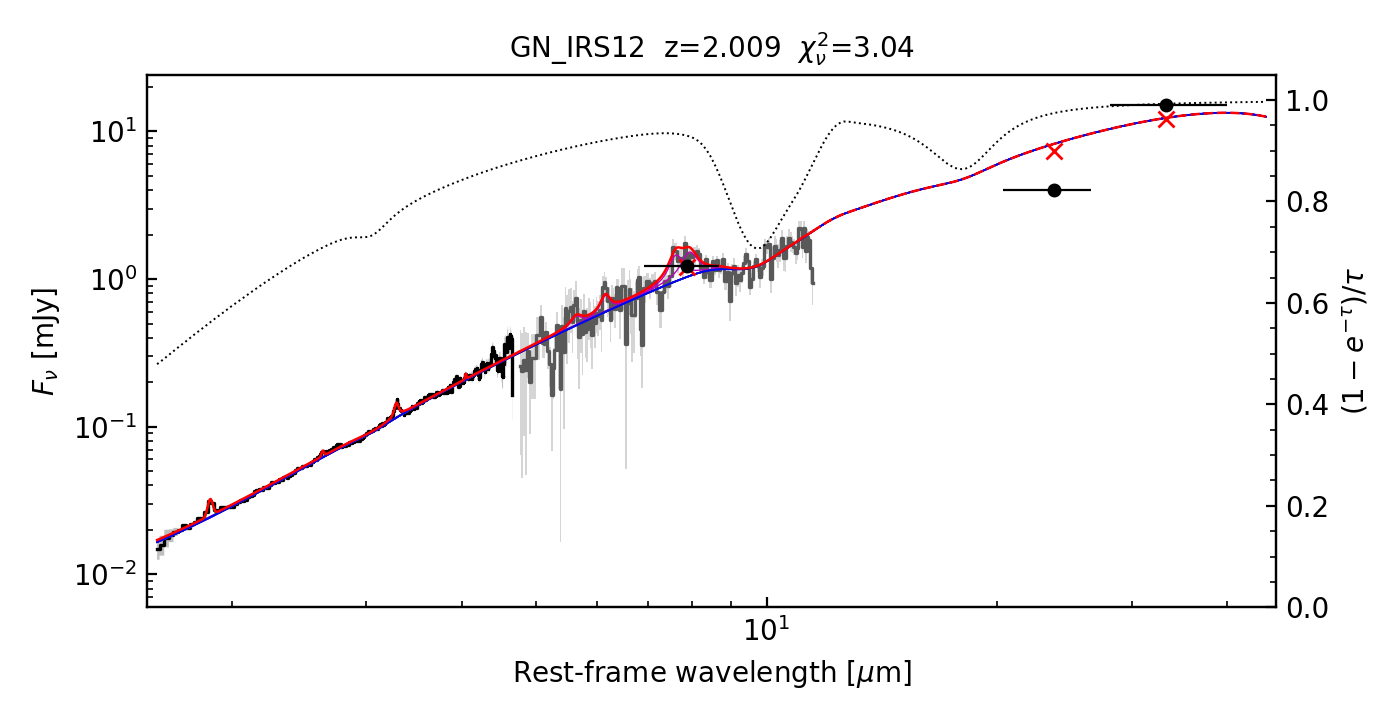}
    \caption{Example mid-infrared decomposition to recover the spectral features measured in our \textit{JWST} MIRI LRS observations combined with \textit{Spitzer} IRS and IR photometry.
    The left panel shows the LRS spectrum of GN-IRS-1 at $z_{\rm spec}=1.4362$. This spectrum exhibits prominent detections of the $3.3\,\mu$m and $3.4\,\mu$m PAH, Br$\alpha$, as well as the stellar CO absorption features at $2.4\,\mu$m and the $3.05\,\mu$m H$_2$O ice absorption feature. The right panel shows GN-IRS-12 at $z_{\rm spec}=2.009$, an AGN-dominated spectrum with Pa$\alpha$ and the $3.3\,\mu$m PAH atop a rising power-law continuum.
    In both panels, the total model is shown in red, with the continuum in blue, and PAH features plus continuum in purple. 
    The dotted black lines in both panels show the inferred extinction model used to dust-correct emission line fluxes assuming a mixed geometry. 
    }
    \label{fig:decomp}
\end{figure*}

For the stellar continuum $I_\nu^*$, we assume a 10 Myr old solar-metallicity stellar population from \texttt{STARBURST99} \citep{Leitherer1999,starburst99}, and let the overall normalization of the template vary. This is needed to fit the CO band heads at rest-frame $2.3\,\mu$m which arise from red supergiants \citep{McGregor1987,Doyon1994}. We test stellar population ages up to 100 Myrs old and find that our assumption of the age of stellar population model has no impact on the inferred PAH luminosities. We model the thermal dust continuum $I_{\rm dust}$ as a power law meant to capture the aggregate emission from cold, warm, and hot dust components, and include \textit{Spitzer} MIPS 24 and 70$\,\mu$m photometry and PACS 70 and 100$\,\mu$m photometry in the fits to anchor this continuum component into the rest-frame mid-infrared. 
Five AGN-dominated galaxies (GN-IRS-12, GN-IRS-19, FLS-IRS-509, FLS-IRS-8226, FLS-IRS-16122) use a cubic polynomial instead which better reproduces the observed shape of continuum in these sources. 
In this way, we perform the necessary spectral decomposition needed to derive accurate PAH line luminosities 
while minimizing the number of free parameters. 
Figure \ref{fig:decomp} shows example model fits for a star-forming galaxy and an obscured AGN, focusing on the LRS spectral region. Extinction-corrected luminosities for the $3.3$, $7.7$, and $11.3\,\mu$m PAH features and the $3.4\,\mu$m aliphatic feature are given in Table \ref{tab:props}, including 3$\sigma$ upper limits in the case of non-detections. 

\begin{figure*}
    \centering
    \includegraphics[width=\textwidth]{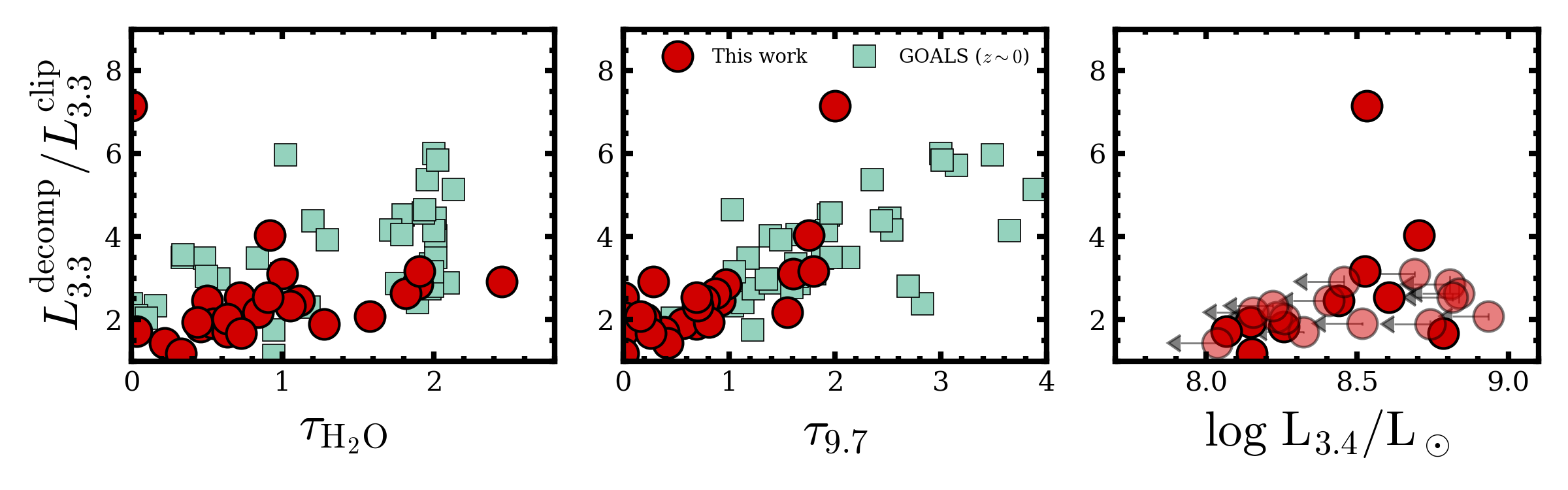}
   \caption{Ratio of the $3.3\,\mu$m PAH line luminosity as measured from the mid-infrared spectral decomposition ($L_{3.3}^{\rm decomp}$) to that of the clipping method ($L_{3.3}^{\rm clip}$) as a function of the $3.05\,\mu$m ice optical depth (\textit{Left} panel), the $9.7\,\mu$m silicate opacity (\textit{Center} panel), and the luminosity of the $3.4\,\mu$m PAH (\textit{Right} panel). We compare our results (red circles) to those for local LIRGs from GOALS (cyan square), which span similar ranges in $L_{3.3}^{\rm decomp}/L_{3.3}^{\rm clip}$ and both optical depths. The difference between the two feature extraction methods most strongly correlates with $\tau_{\rm H_2O}$, $\tau_{9.7}$, and $L_{3.4}$. Jointly modeling water ice absorption, silicate absorption, and $3.4\,\mu$m aliphatic emission is important for inferring accurate $3.3\,\mu$m PAH luminosities at \loglir$\,>11.5$.}
    \label{fig:pah:method}
\end{figure*}

As a consistency check of our spectral decomposition, we also measure the luminosity of the $3.3\,\mu$m PAH directly from the spectrum after subtracting a linear continuum extrapolated from adjacent line-free regions ($L_{3.3}^{\rm clip}$). The continuum anchor points are $3.2$ and $3.4\,\mu$m for the $3.3\,\mu$m PAH, $6$ and $6.6\,\mu$m for the $6.2\,\mu$m PAH, $7.4$ and $8.25\,\mu$m for the $7.7\,\mu$m PAH complex, $8.5$ and $8.9\,\mu$m for the $8.6\,\mu$m PAH complex, and $10.9$ and $11.4\,\mu$m for the $11.2\,\mu$m PAH. In select cases we adjust these anchor points by $<10\%$ to achieve more accurate characterization of the local continuum, especially around the $3.05\,\mu$m water ice feature. Figure \ref{fig:pah:method} shows the ratio of $L_{3.3}^{\rm clip}$ which importantly is not corrected for extinction, to $L_{3.3}$ from the decomposition approach ($L_{3.3}^{\rm decomp}$), as a function of (1) the $3.05\,\mu$m H$_2$O ice opacity which can produce a strong overlapping absorption feature, (2) the $9.7\,\mu$m silicate opacity which sets the total extinction levels, and (3) the luminosity of the adjacent $3.4\,\mu$m aliphatic PAH whose broad Lorentzian wings can contribute to the total flux at $3.3\,\mu$m. We find $L_{3.3}^{\rm decomp}/L_{3.3}^{\rm clip}$ between $1.2$ and $7.2$ (median $2.2$) across our sample. This ratio correlates with the $3.05\,\mu$m H$_2$O ice opacity, the $3.4\,\mu$m aliphatic PAH luminosity and the $\tau_{9.7}$, demonstrating that fitting mid-IR features impacts the inferred $3.3\,\mu$m PAH luminosities. Joint fitting of these features is warranted given the complexity of this spectral region as well as the total extinction. It is reassuring that comparable $\tau$ values and $L_{3.3}^{\rm decomp}/L_{3.3}^{\rm clip}$ are seen in local LIRGs as shown in Figure \ref{fig:pah:method}. We adopt the values from our decomposition method as our fiducial PAH line luminosities but discuss the clipped luminosities throughout the rest of this work as a baseline comparison. 

\begin{figure}
    \centering
    \includegraphics[width=\linewidth]{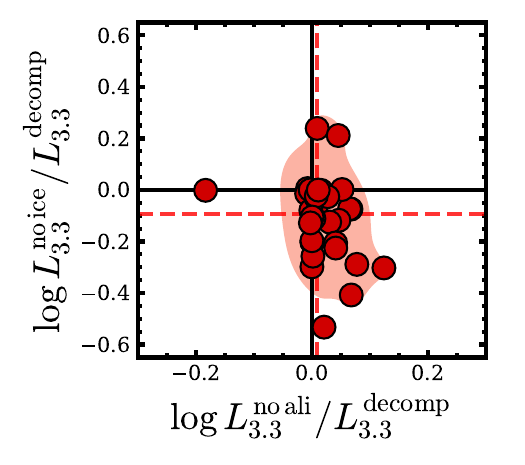}
    \caption{Systematics when measuring the decomposed $3.3\,\mu$m PAH luminosity as shown by turning off the $3.4\,\mu$m aliphatic feature ($L_{3.3}^{\rm no\,ali}$) versus turning off the H$_2$O ice opacity ($L_{3.3}^{\rm no\,ice}$). Red dashed lines show the median change to $L_{3.3}$ along each axis, which is negligible for the $3.4\,\mu$m aliphatic feature  and $\approx0.1$ dex for H$_2$O ice. Given these systematics, we add an additional $0.1$ dex uncertainty to our inferred $3.3\,\mu$m PAH luminosities. Modeling the $3.05\,\mu$m water ice feature is important for getting the adjacent PAH feature luminosity correct. }
    \label{fig:pah:systematics}
\end{figure}

To investigate the systematics on $L_{3.3}^{\rm decomp}$ within our fits, we rerun the decomposition with the $3.4\,\mu$m aliphatic feature turned off and then separately with the H$_2$O ice opacity turned off. The impact of these changes is shown in Figure \ref{fig:pah:systematics}. Choosing whether or not to fit out the $3.4\,\mu$m aliphatic feature does not strongly change the inferred 3.3$\,\mu$m PAH luminosity but fitting the H$_2$O ice opacity does. While the need to fit the ice feature is well motivated by its clear presence in nearly every spectrum with a robust $3.3\,\mu$m PAH detection\footnote{The case for ice absorption is somewhat less obvious in the $z<1$ subset of our sample where H$_2$O ice and the CO band heads are blended due to LRS' low resolving power in the blue end.}, we add an additional $0.1$ dex uncertainty to the luminosity errors to account for these choices in model assumptions. This factor is derived from the median difference between fits with and without the ice opacity. Detecting the continuum to measure H$_2$O opacities is therefore very important for measuring accurate $3.3\,\mu$m PAH luminosities.

\subsection{Comparisons with CAFE}
\begin{figure}[h!]
    \centering
    \includegraphics[width=\linewidth]{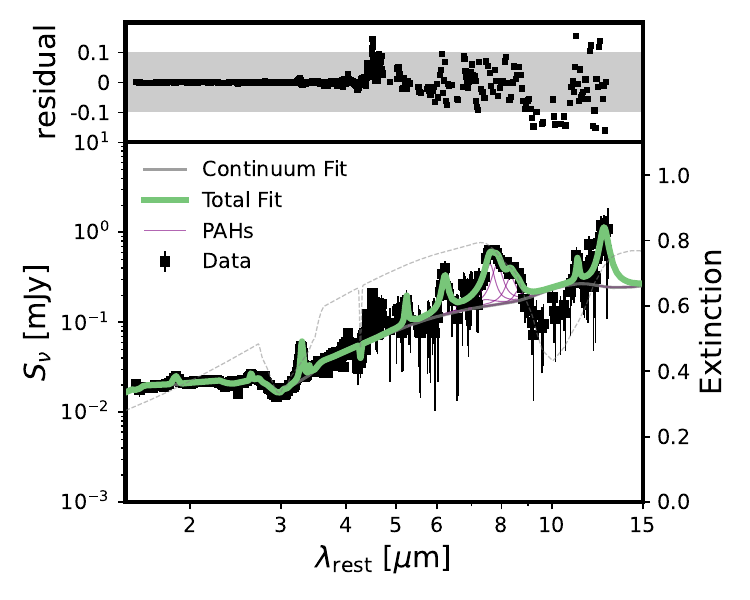}
    \caption{\texttt{CAFE} model fit to the spectrum of FLS-IRS-22530, highlighting the success of \texttt{CAFE} in fitting joint \textit{JWST} MIRI LRS and \textit{Spitzer} IRS spectra of high-redshift dust-obscured star-forming galaxies.}
    \label{fig:cafe:fit}
\end{figure}

As a final benchmark for our simplified mid-infrared decomposition model, we also fit our spectra with \texttt{CAFE} \citep{cafe}. We make a number of simplifying assumptions for the \texttt{CAFE} fits to reduce the number of free parameters down to $30$ and avoid overfitting the observations. We fix the central wavelengths and widths of the PAH features, only include features between $3$ and $12\,\mu$m, and assume that each dust continuum component ($T_{\rm dust}\in[10,1000]$\,K) has the same opacity. With these simplifications, \texttt{CAFE} and our model fit the data equally well. The 16th, 50th, and 84th percentiles for the reduced $\chi^2$ are $5.1^{+8.2}_{-1.8}$ for our model and $6.4^{+9.1}_{-1.5}$ for \texttt{CAFE}. Figure \ref{fig:cafe:fit} shows a representative \texttt{CAFE} fit for one of our observations, demonstrating the success of \texttt{CAFE} in achieving residuals within $10\%$ for $95\%$ of the data between $\lambda_{\rm rest}=1$ and $15\,\mu$m. Throughout the rest of this work we adopt the outputs of our model described in Section \ref{sec:decomp} as our fiducial values for quantities such as PAH line luminosities; however, we use the \texttt{CAFE}-derived numbers as consistency checks to explore model systematics. With appropriate priors, \texttt{CAFE} can be used to fit and interpret \textit{JWST} MIRI LRS spectra.


\begin{figure}
    \centering
    \includegraphics[width=\linewidth]{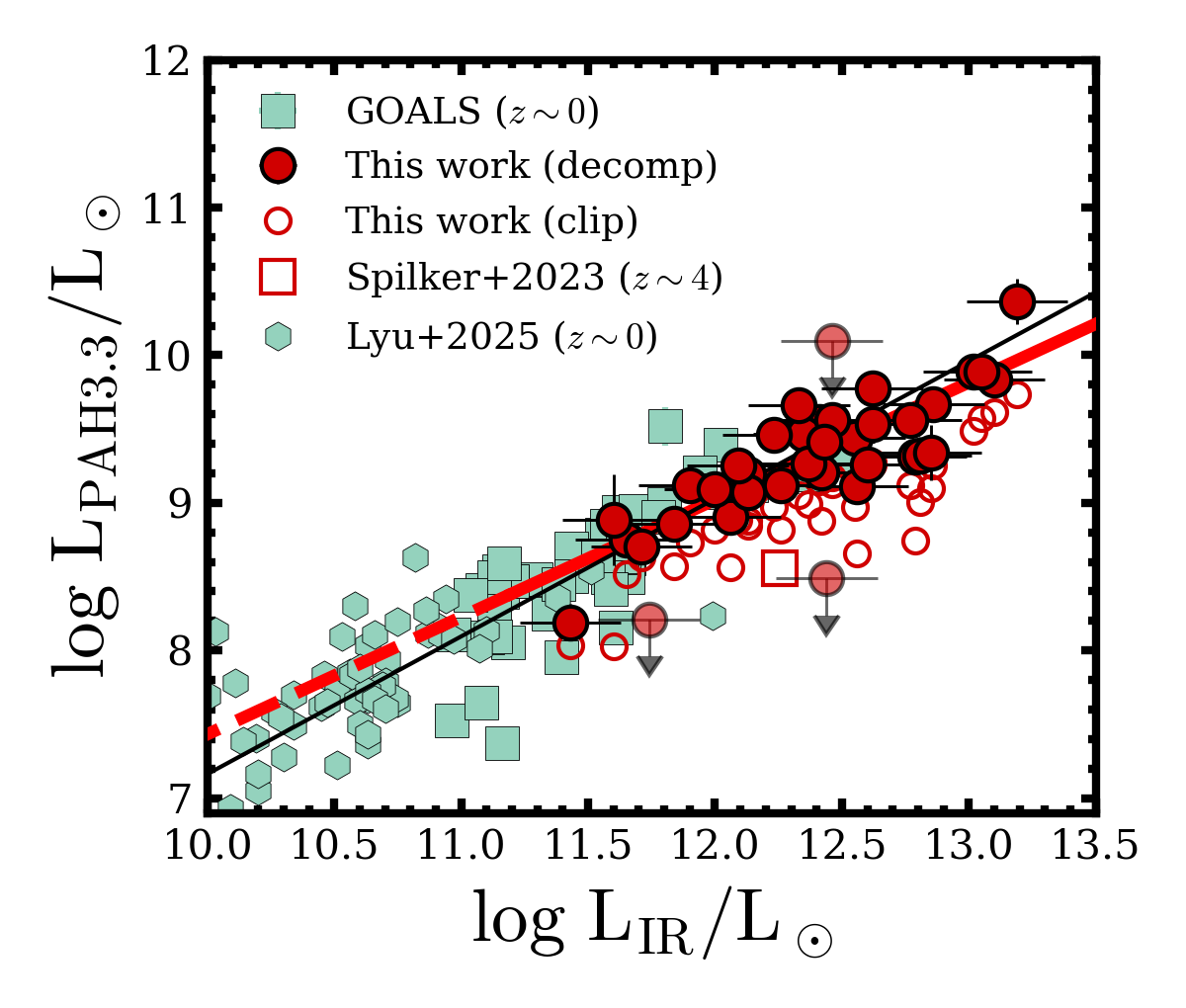}
    \caption{The luminosity of the 3.3$\,\mu$m PAH vs.~total IR luminosity for galaxies in this program (red circles), compared to those at $z\sim0$ from GOALS (cyan squares, corrected for extinction, \citealt{Inami2018,McKinney2021a}) and FRESCO (cyan hexagons, no extinction corrections, \citealt{Lyu2025}). We also show the highest redshift detection of the 3.3$\,\mu$m PAH from \citet{Spilker2023} at $z=4.2248$ in a lensed submillimeter galaxy (red square, no extinction correction). Filled red markers indicate extinction-corrected $L_{3.3}$ for our sample, whereas open red markers correspond to luminosities that are not corrected for extinction. The solid black line shows the best-fit linear trend at $z=0$ to galaxies with \lir$>10^{10}\rm{L_\odot}$, and the red line shows the best-fit linear trend at $z\sim1-2$. The 3.3$\mu$m PAH can comprise up to $\sim0.2$\% of $L_{\rm IR}$ at these redshifts.}
    \label{fig:pah:l33}
\end{figure}

\begin{figure}
    \centering
    \includegraphics[width=\linewidth]{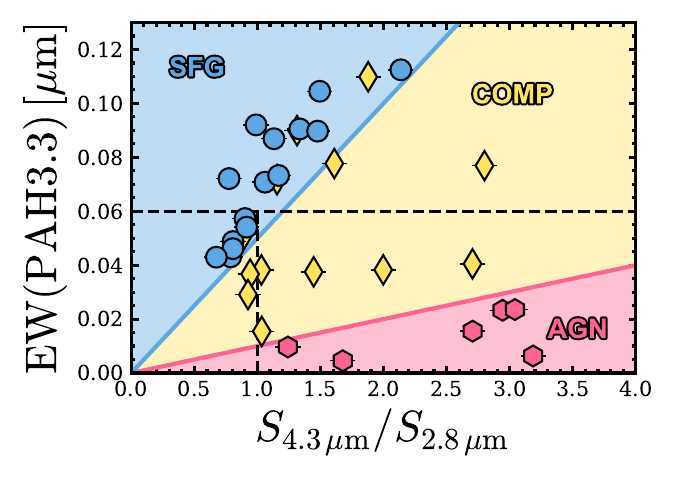}
    \caption{EW of the $3.3\,\mu$m PAH as a function of the near-infrared spectral slope as measured between rest-frame $2.8-4.3\,\mu$m. Galaxies are colored according to their mid-infrared AGN fraction, with star-forming galaxies ($0\%<f_{\rm AGN}<20\%$) shown with blue circles, composites ($20\%<f_{\rm AGN}<50\%$) shown as yellow diamonds, and AGN-dominated galaxies ($f_{\rm AGN}>50\%$) shown with red hexagons. The dashed black lines show the criteria from \cite{Inami2018} used to separate AGN (EW$_{3.3}<0.06\,\mu$m, $S_{4.3}/S_{2.8}>1$), SFGs (EW$_{3.3}>0.06\,\mu$m), and composites (EW$_{3.3}<0.06\,\mu$m, $S_{4.3}/S_{2.8}<1$)  among $z\sim0$ luminous infrared galaxies. We find that mid-infrared AGN at cosmic noon have low $3.3\,\mu$m PAH EWs $<0.02\,\mu$m and that a combination of EW and spectral-slope selection similar but not equal to the $z\sim0$ criteria can be used to separate SFGs and AGN-dominated galaxies at $z\sim1-2$. 
    }
    \label{fig:pah:ew33}
\end{figure}

\begin{figure}
    \centering
    \includegraphics[width=\linewidth]{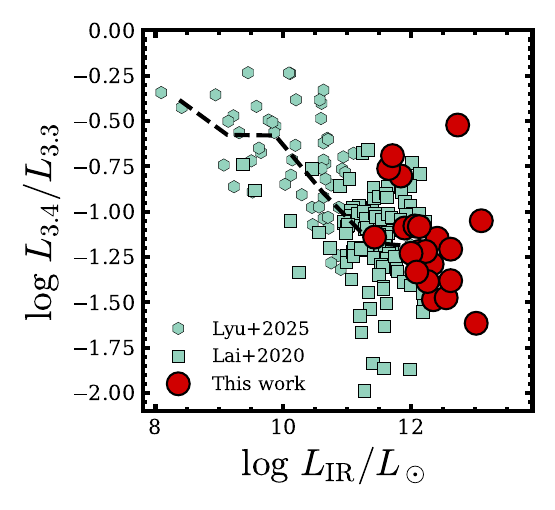}
    \caption{Luminosity ratio of the $3.4\,\mu$m aliphatic feature to the $3.3\,\mu$m PAH in our $z\sim1-2$ sample (red) compared to $z=0-0.5$ galaxies from \cite{Lyu2025} and \cite{Lai2020} (cyan) as a function of total infrared luminosity. This ratio probes the aliphatic fraction which at $z\sim1-2$ is consistent with the high \lir\ subset of galaxies from the low$-z$ samples. The dashed black line shows the moving average among the low$-z$ galaxies and the shaded red region encases 84\% of $z\sim1-2$ galaxies when inferring $L_{3.4}/L_{3.3}$ using the same methods as were done for \cite{Lai2020}. 
    }
    \label{fig:pah:ali}
\end{figure}

\section{Results\label{sec:results}}

In this section, we present the results from our decomposition of the joint \textit{JWST} MIRI LRS and \textit{Spitzer} IRS spectra. In the scope of this overview paper, we focus on the total luminosity and EW of the 3.3$\,\mu$m PAH feature which we are statistically measuring for the first time at high redshifts. Throughout this section, we compare to the luminosity of this feature measured in nearby galaxy samples such as GOALS \citep{Armus2009} and FRESCO \citep{Oesch2023}, as well as single-object studies at $z\sim4$ \citep{Spilker2023}. Where possible, we compare to measurements made using similar assumptions to the ones we adopt for spectral decomposition as well as when measuring total infrared luminosities. For instance, \lir\ in GOALS galaxies are also integrated between $8$ and $1000\,\mu$m \citep{DiazSantos2013,DiazSantos2017} consistent with how \lir\ is measured in our $z\sim1-2$ sample \citep{Kirkpatrick2017}.

LIRGs from the GOALS sample are the closest local analogs to $z\sim1-2$ dusty, star-forming galaxies; however, they tend to fall above the star formation main sequence by factors of $6-70$ \-- firmly in the locus of ``starburst'' galaxies \citep{U2012,Kirkpatrick2017}. The galaxies in our high-redshift sample have SFRs $\sim2-20\times$ higher than the main-sequence value for their stellar masses \citep{Kirkpatrick2017}. A minority ($\sim25\%$) are within the main-sequence scatter for star-forming galaxies with $\log\,M_*/M_\odot>10.5$. While the rest of the sample galaxies are categorically starbursts (${\rm sSFR/sSFR_{MS}>3}$, \citealt{Speagle2014}), they do not reach the same range of distance above the main sequence seen for local LIRGs. Therefore, any evolution we infer by comparing our galaxies to GOALS is also capturing the different positions along the star formation main sequence that both samples span. Nevertheless, both samples represent the extremes of dust-obscured star formation for their respective epochs.

\subsection{The Luminosity of the 3.3$\mu$m PAH}
In high-redshift galaxies the PAH emission features at $6.2$, $7.7$, and $11.3\,\mu$m are independently very bright and can each comprise $1-4\%$ of \lir\ \citep{Pope2008}. Figure \ref{fig:pah:l33} shows the relation between $L_{3.3}$ and \lir\ from our LRS spectra, using luminosities corrected for extinction ($L_{3.3}^{\rm decomp}$) and not ($L_{3.3}^{\rm clip}$). Despite being one of the faintest PAH features, the $3.3\,\mu$m PAH luminosities observed in our sample are typically $0.06-0.15\%$ of \lir\ and can be as bright as $\sim10^{10.4}\,\rm{L_\odot}$. We find a best-fit linear relation between $L_{3.3}$ and \lir\ for $z\sim1-2$ galaxies of
\begin{equation}
    \log\frac{L_{\rm 3.3}}{L_\odot} = 0.79\,(\pm0.05)\times \log\frac{L_{\rm IR}}{L_\odot}-0.5\,(\pm0.5)
\end{equation}
This relation lies on the relation from low$-z$ surveys of the $3.3\,\mu$m PAH from \citet{Inami2018} (local LIRGs) and \cite{Lyu2025} (normal star-forming galaxies from $z=0.2-0.5$). 
We note that the high-redshift $L_{3.3}$ luminosities not corrected for dust extinction fall $0.37$ dex below the low$-z$ relation on average. 
\cite{Spilker2023} report the detection of the $3.3\,\mu$m PAH at $z=4.2248$ in a lensed submillimeter galaxy with  $L_{3.3}$/\lir$\,\sim0.01\%$ making the line $\sim6\times$ less luminous than what is observed in our $z\sim1-2$ sample; however, we note that \cite{Spilker2023} do not correct for any dust extinction which would imply a higher value of $L_{3.3}$, possibly in agreement with our extinction-corrected luminosities. 


\subsection{The Equivalent Width of the 3.3$\mu$m PAH}

EWs of PAH features are commonly used to discriminate between star-forming galaxies and AGN on the basis of the feature strength powered by star formation competing with an AGN-driven continuum \citep{Moorwood1986,Imanishi2006,Spoon2007}. In Figure \ref{fig:pah:ew33}, we compare the EW of the 3.3$\mu$m PAH, as measured directly from the spectrum using a spline-like method, as a function of the LRS spectral slope calculated as the slope between rest-frame $2.8-4.3\,\mu$m. In addition, 3.3$\mu$m PAH EWs are included in Table \ref{tab:props}. Galaxies are separated by mid-infrared AGN fraction in a manner different from the classification scheme used earlier. Here, we identify star-forming galaxies as having $0\%<f_{\rm AGN}<20\%$, composites with  $20\%<f_{\rm AGN}<50\%$, and AGN-dominated galaxies with $f_{\rm AGN}>50\%$. This empirical diagnostic follows from \cite{Inami2018} who use $3.3\,\mu$m PAH EWs and near-infrared slopes to separate star-forming galaxies and AGN using \textit{AKARI} $2-5\,\mu$m spectra of nearby luminous infrared galaxies. \cite{Inami2018} found that at $z\sim0$ the $3.3\,\mu$m PAH EWs of $0.06\,\mu$m or less generally selected the AGN-dominated systems. In our sample, we find that dust-obscured $z\sim1-2$ AGN typically have $3.3\,\mu$m PAH EWs less than $0.02\,\mu$m. The star-forming galaxies have $3.3\,\mu$m PAH EWs between $0.04-0.12\,\mu$m. While there is not a clear correlation between the near-infrared spectral slope and $3.3\,\mu$m PAH EW, the following criterion select all the AGN and SFGs, respectively: 
\begin{equation}
    {\rm SFG} : {\rm EW(3.3)/\,\mu m} > 0.05\times(S_{4.3\mu m}/S_{2.8\mu m})  
\end{equation}
\vspace{-15pt}
\begin{equation}
    {\rm AGN} : {\rm EW(3.3)/\,\mu m} < 0.01\times(S_{4.3\mu m}/S_{2.8\mu m})  
\end{equation}
where $S_{4.3\mu m}$ and $S_{2.8\mu m}$ correspond to the observed flux densities at rest-frame wavelengths of $4.3\,\mu$m and $2.8\,\mu$m, respectively. These criteria may be used to identify dust-obscured AGN in high-redshift galaxies using LRS spectra alone. 


\subsection{Aliphatic PAHs in High-redshift Galaxies}
The detection of the $3.4\,\mu$m feature in our spectra provides a handle on the aliphatic\footnote{Carbon chains, non-aromatic.} fraction ($f_{\rm ali}$) of grains at $z\sim1-2$. Recently, \cite{Lyu2025} tested $f_{\rm ali}$ using the $3.4/3.3\,\mu$m aliphatic-to-aromatic ratio as an empirical proxy, finding the most significant correlations against \lir\ and the SFR. Figure \ref{fig:pah:ali} shows the $3.4/3.3\,\mu$m aliphatic-to-aromatic ratio as a function of \lir\ for this work compared against low$-z$ galaxies from \cite{Lyu2025} and \cite{Lai2020}. At high redshift, we find $3.4/3.3\,\mu$m ratios generally consistent with the most luminous local infrared galaxies. The range of $z\sim1-2$ galaxies along $3.4/3.3\,\mu$m versus \lir\ does not depend on the PAH luminosity extraction method. 

\section{Discussion\label{sec:disc}}

\subsection{Tracing Dust-obscured Star Formation at $z>1$ with the $3.3\,\mu$m PAH}
\begin{figure}
    \centering
    \includegraphics[width=\linewidth]{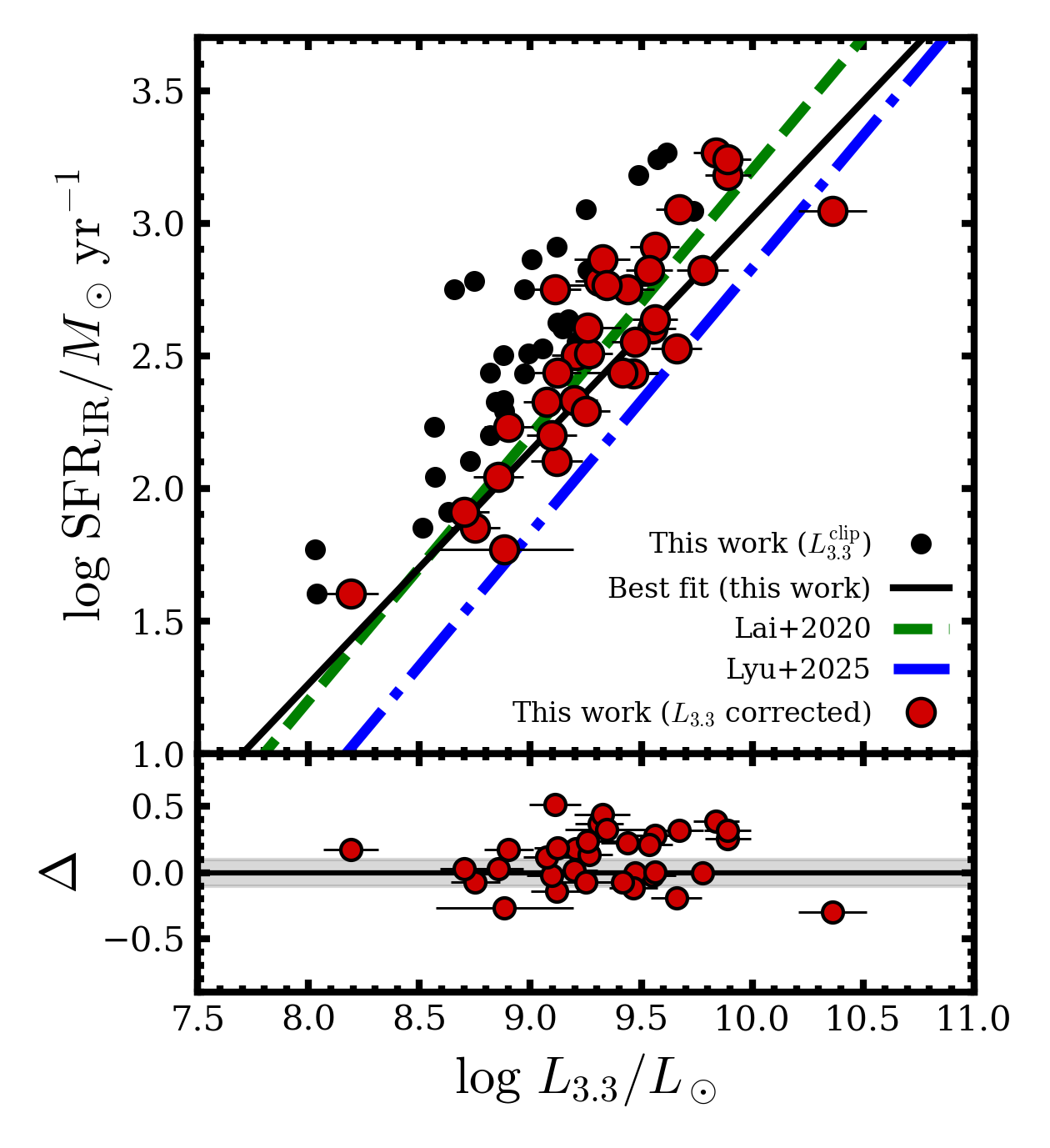}
    \caption{The correlation between IR star-formation rate (${\rm SFR_{IR}}$) and $3.3\,\mu$m PAH luminosity at $z\sim0.8-2$ in massive, metal-enriched galaxies with ${\rm SFR_{IR}}=30-1600\,M_\odot\,{\rm yr^{-1}}$. We show the trend for $3.3\,\mu$m PAH luminosities corrected for extinction (solid red, black solid line). $L_{3.3}$ with no extinction corrections applied are shown as black circles. For comparison, we also show the SFR-$L_{3.3}$ calibrations from \cite{Lai2020} which span $2<{\rm SFR/}M_\odot \,{\rm yr^{1}}<320$ at $z\sim0$ (green), and from \cite{Lyu2025} at $z=0.2-0.5$ and ${\rm SFR/}M_\odot \,{\rm yr^{1}}<20$ (blue). Our extinction-corrected star-formation rate calibration is accurate to $\pm0.2$ dex between $\log L_{3.3}/L_\odot\sim8.5-10.5$, as shown in the lower panel.}
    \label{fig:pah:sfr}
\end{figure}

PAH emission features are typically ascribed to infrared fluorescence following the absorption of far-ultraviolet photons, and are therefore good obscured SFR indicators \citep{Peeters2004,Brandl2006,Smith2007,Pope2008}. To calibrate the $3.3\,\mu$m PAH as an SFR indicator at high redshift, we begin by using Equation (5) from \cite{Kirkpatrick2015} to convert mid-infrared AGN fractions to a total infrared AGN fraction, $f_{\rm AGN,tot}$. Next, we correct \lir\ for the contribution from AGN heating of dust using $L_{\rm IR}^{\rm SF}=(1-f_{\rm AGN,tot})\times L_{\rm IR}$. Finally, we estimate the IR SFR using ${\rm SFR_{IR}\,/\,}M_\odot\,{\rm yr^{-1}}=1.59\times10^{-10}\,L_{\rm IR}^{\rm SF}$ from \cite{Murphy2011}. Figure \ref{fig:pah:sfr} shows the tight correlation between the $3.3\,\mu$m PAH luminosity and IR SFR from galaxies in our sample. The best-fit linear relation is given by 
\begin{equation}
    \log\,\frac{\rm SFR_{IR}}{M_\odot\,{\rm yr^{-1}}} = -5.77\,(\pm0.84) + 0.88\,(\pm0.09)\log \frac{L_{3.3}}{L_\odot}
\end{equation}
and exhibits a $\sim10\%$ scatter. 

Our SFR-$L_{3.3}$ relation is within the scatter of the $z<1$ calibrations \cite{Lai2020}\footnote{We compare to the mixed geometry calibration of \cite{Lai2020} for consistency with our assumptions when modeling extinction.}, so the relation is comparable across $z\sim0-2$. 
We note that \lir\ for each comparison sample is integrated over the same range as our measurements ($8-1000\,\mu$m) and is calculated empirically from \textit{Spitzer} and \textit{Herschel} photometry, except for \cite{Lyu2025} who scale a far-infrared template to $24\,\mu$m MIPS photometry to estimate \lir. This may contribute to their systematic offset of $0.27$ dex from our work and \cite{Lai2020}. 

In future observations with \textit{JWST} the $3.3\,\mu$m PAH may be detected in spectra that do not recover the adjacent continuum. As discussed previously, correcting for $3.05\,\mu$m ice absorption is important for deriving accurate, extinction-corrected $3.3\,\mu$m PAH luminosities. However, the uncorrected feature luminosities may still provide a first-order estimate of the obscured SFR. In this case, we note that the median offset between the corrected and not corrected luminosities is $\langle\log  L_{3.3}^{\rm decomp}/L_{3.3}^{\rm clip} \rangle =0.33^{+0.16}_{-0.09}$. 
This correction factor should be used with caution for $z>1$ galaxies having $\log\,M_*/M_\odot\gtrsim10$ and ${\rm SFR}>30\,M_\odot\,{\rm yr^{-1}}$.

\subsection{PAH Properties in High-redshift Galaxies}

The intensity ratios between various PAH lines are sensitive to the underlying size and charge distributions of PAH themselves \citep{Draine2001,Draine2007}, as well as their excitation conditions \citep{Maragkoudakis2020}. In local luminous infrared galaxies, these PAH ratios are largely invariant across star-forming galaxies and AGN \citep{Stierwalt2014}. There is some evidence that high-mass, metal-rich galaxies at $z\sim2$ have PAH ratios consistent with those found at $z\sim0$ \citep{McKinney2020}, suggesting comparable PAH excitation conditions and/or grain properties; however, the $3.3\,\mu$m PAH has never been statistically studied at high redshift alongside the longer wavelength PAHs. 

\begin{figure}
    \centering
    \includegraphics[width=\linewidth]{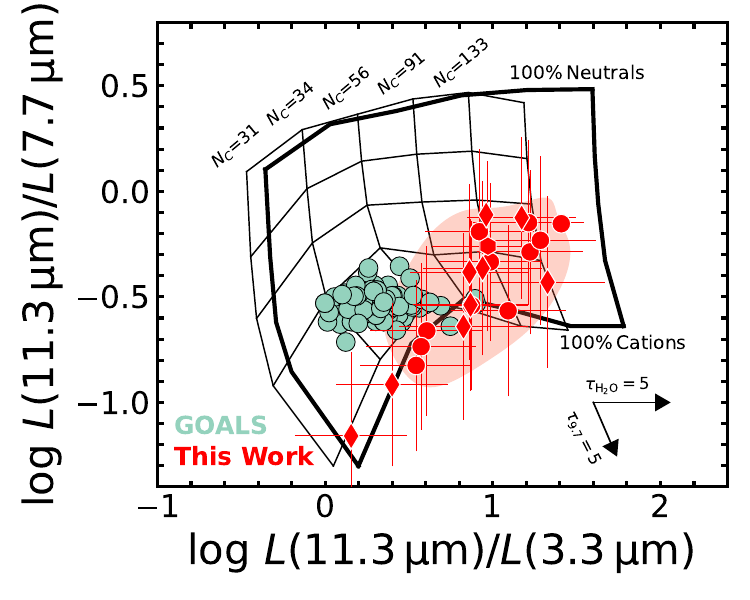}
    \caption{Grain size and charge diagnostics using ratios of the $3.3\,\mu$m, $7.7\mu$m, and $11.3\mu $m PAH features. The thin black grid shows the dependence of these ratios on the size and neutral fraction of PAHs from \citet{Maragkoudakis2020} assuming a standard ISRF density (${\langle h \nu\rangle}\sim1$ eV). The thick black outline shows how the grid shifts when assuming single photon heating with photon energies of 6 eV. Our sample is shown in red, with symbols following the AGN fraction classifications of Figure \ref{fig:pah:ew33}. The red region corresponds to our measurements using \texttt{CAFE}. 
    Extinction-corrected PAH line ratios for local LIRGs from GOALS are shown in turquoise \citep{Stierwalt2013,Stierwalt2014,Inami2018,McKinney2021a}. In the lower right corner, we show how the PAH line ratios change when increasing just the 9.7$\mu$m silicate optical depth ($\tau_{9.7}$) or just the $3.05\mu$m H$_2$O optical depth ($\tau_{\rm H_20}$), keeping the strength of the lines fixed and assuming a mixed geometry. Dust-obscured star-forming galaxies at $z\sim1-2$ have larger PAHs on average than $z\sim0$ LIRGs. This result holds even if we use the clipped PAH line luminosities (see Section \ref{sec:decomp}) or \texttt{CAFE}-derived luminosities, and is therefore robust against the mid-infrared decomposition method.}
    \label{fig:pah:ratios}
\end{figure}

Figure \ref{fig:pah:ratios} shows the ratio of the $11.3\mu$m PAH luminosity to that of the $7.7\mu$m and $3.3\mu$m PAHs. For the PAH lines that are accessible at $z>0$, this particular combination provides the best joint constraint on the size of PAH grains and their neutral fraction \citep{Maragkoudakis2020}. Details on how these model grids were calculated can be found in \cite{Maragkoudakis2020}. As a brief summary of the work therein, emergent PAH band fluxes were calculated using theoretical absorption spectra exposed to various radiation fields for PAHs with a number of C atoms ($N_C$) above 20. Boundaries labeled 100\% neutral and 100\% cations on Figure \ref{fig:pah:ratios} correspond to feature ratios predicted for pure neutral or cationic mixtures. For comparison we show the model grid computed for the ISRF model of \cite{Mathis1983} with mean photon energy $\sim1$ eV. To illustrate the direction in which the model grids change when increasing the strength of the incident radiation field, we also show the boundaries for a model with $\langle h\nu\rangle=6$ eV. While the absolute position of model grids shown on figures such as Figure \ref{fig:pah:ratios} show some variability between models, the directional dependence of PAH feature ratios with $N_C$ and neutral fraction is generally preserved \citep{Draine2001,Maragkoudakis2020}.

Dust-obscured galaxies at high redshift show an offset from local LIRGs with respect to the $11.3/3.3\,\mu$m PAH ratio. In our sample we find a median $\log\,L_{11.3}/L_{3.3}=0.96^{+0.31}_{-0.39}$, which is $3\times$ larger than that of local LIRGs: $\log\,L_{11.3}/L_{3.3}=0.32^{+0.16}_{-0.16}$. We do not find as strong of an offset for the $11.3/7.7\,\mu$m ratio, for which we measure a median $\log\,L_{11.3}/L_{7.7}=-0.41^{+0.24}_{-0.30}$ in our $z\sim1-2$ galaxies as compared to $\log\,L_{11.3}/L_{7.7}=-0.53^{+0.05}_{-0.05}$ for local LIRGs. Explaining the $11.3/3.3\,\mu$m ratio offset with extinction alone would require $3.05\,\mu$m ice opacities of 5 or greater assuming a mixed geometry, which is not recovered by our decomposition method where $\tau_{3.05}<2.5$ (Figure \ref{fig:pah:method}). This offset between local and high$-z$ galaxies persists when using PAH line luminosities extracted with \texttt{CAFE} in the exact same manner as has been done in GOALS and also if we use the clipped PAH line luminosities that make no dust extinction correction. Based on the \cite{Maragkoudakis2020} models, we expect this offset to arise from a systematic shift toward larger PAH grain sizes, or changes to the average ISRF that the PAHs are subject to, or a combination of both. 

In the local Universe, the high $11.3/3.3\,\mu$m PAH ratios we observe in our sample are only found in regions where the ISM is exposed to very strong ultraviolet fields \citep{Lai2022,Lai2023,Lai2025}. 
The fact that high $11.3/3.3\,\mu$m PAH ratios are more common at $z>1$ could therefore  be consistent with the larger ionization parameters and higher cold gas excitation conditions found in high-redshift galaxies \citep{Boogaard2020,Hsiao2022,Shen2025}, as well as the on average larger \lir\ range spanned by our $z\sim1-2$ sample relative to local LIRGs from GOALS. 
As indicated in Figure \ref{fig:pah:ratios}, the $11.3/3.3\,\mu$m ratios are not biased by mid-infrared AGN fraction. AGN are not systematically influencing these line ratios, which are instead likely set by the local ISM conditions driven by star formation. It is possible that our galaxy-integrated measurements are not sensitive to changes to the PAHs close to an AGN because of spatial blending, in which case the observed PAH line ratios would be arising from star formation in the AGN-host galaxy. 

Another way to achieve high $11.3/3.3\,\mu$m PAH ratios would be to lower the mean photon energy incident on the PAHs (see Figure \ref{fig:pah:ratios}), but this is largely inconsistent with the aforementioned redshift trends in radiation field properties and \lir\ at $z>1$. 
Assuming the $11.3/3.3\,\mu$m PAH ratios represent a luminosity-weighted average over star-forming gas, \citeauthor{Maragkoudakis2020}'s (\citeyear{Maragkoudakis2020}) models suggest that star-forming regions in our sample may host larger PAH grains on average than what is found in local galaxies. Larger PAH grain sizes would be consistent with high ISM densities, corroborated by the water ice detected in our spectra which require 
$n>10^3\,{\rm cm^{-3}}$ and T$<90$ K above which the water ice sublimates to gas phase H$_2$O \citep{Spoon2003,Boogert2015,Sajina2025}. The ultraviolet radiation fields penetrating into these regions of the ISM may still be strong, as evidenced by low $3.4/3.3\,\mu$m ratios consistent with radiative stripping of the aliphatic side groups \citep{Lyu2025}. At such high densities and low temperatures, PAH grain coagulation can dominate over destruction channels like shattering \citep{Seok2014}, which could lead to further enhancement in the mean size of PAHs in addition to photodestruction of the smaller molecules. 

A shift in the PAH grain size distribution toward larger grains carries important implications for the photoelectric heating effect whereby PAHs absorb ultraviolet photons from hot, young stars and transmit that energy into the gas. Photoelectric heating is the dominant heating mode of the cold and warm neutral medium \citep{Tielens1985}. Local luminous infrared galaxies have far-infrared and PAH line ratios consistent with low photoelectric heating efficiencies, with a strong dependence on the infrared surface density \citep{McKinney2021a}. This means that at high dust-obscured SFR surface densities, PAHs are transmitting less energy into the ISM per absorbed ultraviolet photon. Consequently, more energy per photon is transmitted to PAHs. Massive, metal-enriched galaxies at $z\sim1-2$ have high PAH-to-total infrared luminosity ratios \citep{Pope2008,Pope2013}, suggesting that they too may have low photoelectric efficiencies. By preventing a higher fraction (per photon) of the stellar energy from ending up in the gas, low photoelectric heating efficiencies may help support star formation sustained at high rates ($>100\,{\rm M_\odot\,yr^{-1}}$) over the few hundred Myr duty cycle of a dusty starburst \citep{Swinbank2014,Dudzeviciute2020,Sun2021}. An empirical consequence of this effect at play would be low far-infrared fine-structure emission line luminosities (e.g., [C\,{\sc II}] at $\lambda_{\rm rest}=157.7\,\mu$m) relative to the PAHs, for which there is some evidence in massive, metal-rich, mid-infrared--bright $z\sim2$ galaxies selected from the same parent sample as the targets of this work \citep{McKinney2020}. 


\subsection{PAHs in AGN at Cosmic Noon}
We detect PAHs in emission and aliphatic features in absorption from galaxies with near- through mid-infrared spectra dominated by the hot dust emission arising from an AGN torus. By temperature arguments alone, the $3.3\,\mu$m feature would be the least likely PAH to be diluted by hot dust from the AGN torus, which suggests that this effect is the main reason for the absence of PAH lines at longer wavelengths in AGN hosts. The relatively narrower width of the $3.3\,\mu$m PAH profile also allows it to stand out against the underlying continuum. PAHs are likely still present in AGN-host galaxies, we just cannot see the features above the high continuum levels.  

In at least three of our target AGN we detect the $3.4\,\mu$m aliphatic feature in absorption which arises from C-H bonds along the line-of-sight toward a hot, background continuum source \citep{Dartois2007,Gunay2018}, which in this case is likely the dusty torus around the central supermassive black hole. Given the strength of the $3.4\,\mu$m absorption, and the lack of similar profiles detected in star-forming galaxies, one  possible scenario is that the absorbers are present in the torus itself. This would imply significant amounts of shielding to prevent the aliphatic side groups from being radiatively stripped, which they do so readily in star-forming gas \citep{Lyu2025}. However, it is also possible that the absorbers are far from the continuum source and are only apparent because of the very strong continuum and high dust column density in the host galaxy.

\section{Conclusion\label{sec:conc}}
In this work we present the first statistically significant spectroscopic survey of $z\sim1-2$ luminous infrared galaxies making use of \textit{JWST} LRS. Our 37 targets have archival \textit{Spitzer} observations, which we make use of to achieve complete $\lambda_{\rm obs}\sim5-30\,\mu$m spectral coverage. Using these data, we present the first statistical analysis of the $3.3\,\mu$m PAH feature at $z>0.5$. Our main conclusions are as follows: 
\begin{enumerate}
    \item The $3.3\,\mu$m PAH is detected in 34 of 37 galaxies, with the nondetections found among our targets with the highest contribution to their mid-infrared spectrum by an AGN. The luminosity of this line, which is faintest of the PAHs, can reach up to $0.2\%$ of the total infrared luminosity. Comparable to $z\sim0$ studies, a combination of the $3.3\,\mu$m PAH EW and near-infrared spectral slope can distinguish between pure dust-obscured star-forming galaxies and AGN. 
    \item H$_2$O ice absorption is prominent in nearly every star-forming galaxy with a strong detection of the $3.3\,\mu$m PAH and must be modeled jointly to derive dust-corrected PAH line luminosities. This absorption feature is not apparent in the strongest mid-infrared AGN. Instead, the $3.4\,\mu$m aliphatic absorption feature is found in these sources.  
    \item In addition to the $3.3\,\mu$m PAH, we detect atomic emission lines such as Br$\alpha,\gamma$, Pa$\alpha$ and Fe\,II, as well as molecular features in emission, such as PAHs at $3.4-6.2\,\mu$m, and CO gas and CO$_2$ ice in absorption.
    \item We find a tight correlation between the $3.3\,\mu$m PAH luminosity and the dust-obscured SFR for galaxies with ${\rm SFR_{IR}}=30-1600\,M_\odot\,{\rm yr^{-1}}$ that is consistent with the low$-z$ relation from \cite{Lai2020}. The relation has a small $0.12$ dex intrinsic scatter spanning $\log L_{3.3}/L_\odot=8.7-9.8$.
    \item Combining the $3.3\,\mu$m PAH with PAH lines at $7.7\,\mu$m and $11.3\,\mu$m detected by \textit{Spitzer}, we find that $z\sim1-2$ galaxies have $11.3/3.3\,\mu$m PAH line ratios $\approx3\times$ higher than those found in $z\sim0$ luminous infrared galaxies. The increase in this PAH line ratio is attributed to a shift in the PAH grain size distribution toward larger grains, and to a lesser extent, softer incident ultraviolet radiation fields. We find no dependence for the PAH line ratios on the mid-infrared AGN fraction. 
    \item The low aliphatic fraction of PAHs, as probed by the $3.4/3.3\,\mu$m aliphatic-to-PAH ratio, is consistent with the extrapolation of local trends toward high infrared luminosities. 
\end{enumerate}
This work represents the first large extragalactic survey using \textit{JWST} MIRI's LRS instrument. We developed spectral extraction methods to mitigate limiting background effects, revealing exquisite $5-14\,\mu$m spectra rich in emission and absorption features. In this first analysis of the data we focus on the $3.3\,\mu$m PAH emission line, increasing the number of detections beyond $z\sim0.5$ by an order of magnitude. Looking forward, the $3.3\,\mu$m PAH will make a promising diagnostic of the dust-obscured SFR at high redshifts (Eleazer et al. 2026, in preparation), and other features such as the CO and H$_2$O ices can be modeled to constrain the radiative and thermal conditions of the gas (Sajina et al. 2026, in preparation). Future mid-infrared studies with existing facilities such as \textit{JWST} or mission concepts like \textit{PRIMA}\footnote{\url{https://prima.ipac.caltech.edu}} are needed to leverage these diagnostic tools for larger and more heterogeneous galaxy samples.

{\vspace{10pt}\small
J.M. acknowledges S. Kendrew, A. Petric, and the whole MIRI Instrument Science team who supported this project. 
J.M. acknowledges the invaluable labor of the maintenance and clerical staff at our institutions, whose contributions make our scientific discoveries a reality. Authors from UT Austin acknowledge that UT is an institution that sits on indigenous land. The Tonkawa lived in central Texas, and the Comanche and Apache moved through this area. We pay our respects to all the American Indian and Indigenous Peoples and communities who have been or have become a part of these lands and territories in Texas. We are grateful to be able to live, work, collaborate, and learn on this piece of Turtle Island. 

J.M. thanks NASA and acknowledges support through the Hubble Fellowship Program, awarded by the Space Telescope Science Institute, which is operated by the Association of Universities for Research in Astronomy, Inc., for NASA, under contract NAS5-26555. This work is based [in part] on observations made with the NASA/ESA/CSA James Webb Space Telescope. The data were obtained from the Mikulski Archive for Space Telescopes at the Space Telescope Science Institute, which is operated by the Association of Universities for Research in Astronomy, Inc., under NASA contract NAS 5-03127 for JWST. These observations are associated with program \#3224. Support for program \#3224 was provided by NASA through a grant from the Space Telescope Science Institute, which is operated by the Association of Universities for Research in Astronomy, Inc., under NASA contract NAS 5-03127.
}

\begin{deluxetable*}{lcccccccc}
\tablecaption{Target Properties and PAH Luminosities (corrected)\label{tab:props}}
\tablehead{\colhead{ID} & \colhead{$z_{\rm spec}$} & \colhead{$\log L_{\rm IR}/L_\odot$} & \colhead{$f_{\rm AGN}$/\%} & \colhead{EW$_{3.3}$/$\mu$m} & \colhead{$\log L_{3.3}/L_\odot$} & \colhead{$\log L_{3.4}/L_\odot$} & \colhead{$\log L_{7.7}/L_\odot$} & \colhead{$\log L_{11.3}/L_\odot$} \\
\colhead{(1)} & \colhead{(2)} & \colhead{(3)} & \colhead{(4)} & \colhead{(5)} & \colhead{(6)} & \colhead{(7)} & \colhead{(8)} & \colhead{(9)}}
\startdata
GN-IRS-1 & 1.436 & 12.4 & 19 $\pm$ 5 & $0.087 \pm 0.002$ & $9.55 \pm 0.11$ & $8.60 \pm 0.05$ & $10.76 \pm 0.03$ & $10.34 \pm 0.06$ \\
GN-IRS-2 & 1.243 & 12.1 & 0 $\pm$ 5 & $0.057 \pm 0.001$ & $9.20 \pm 0.11$ & $<8.23$ & $10.68 \pm 0.04$ & $10.34 \pm 0.03$ \\
GN-IRS-3 & 1.515 & 12.4 & 0 $\pm$ 5 & $0.091 \pm 0.002$ & $9.47 \pm 0.10$ & $8.26 \pm 0.12$ & $10.76 \pm 0.08$ & $10.06 \pm 0.09$ \\
GN-IRS-4 & 1.264 & 12.3 & 16 $\pm$ 3 & $0.090 \pm 0.001$ & $9.66 \pm 0.11$ & $8.70 \pm 0.04$ & $10.68 \pm 0.02$ & $10.06 \pm 0.12$ \\
GN-IRS-5 & 1.15 & 12.2 & 0 $\pm$ 5 & $0.071 \pm 0.001$ & $9.46 \pm 0.11$ & $<8.69$ & $10.69 \pm 0.07$ & $10.19 \pm 0.06$ \\
GN-IRS-6 & 1.009 & 11.9 & 0 $\pm$ 5 & $0.049 \pm 0.001$ & $9.12 \pm 0.12$ & $<8.41$ & $10.60 \pm 0.07$ & $<10.76$ \\
GN-IRS-7 & 1.998 & 12.6 & 0 $\pm$ 5 & $0.112 \pm 0.004$ & $9.44 \pm 0.12$ & $<8.46$ & $<10.75$ & \nodata \\
GN-IRS-11 & 1.989 & 12.6 & 23 $\pm$ 7 & $0.040 \pm 0.002$ & $9.11 \pm 0.12$ & $<8.81$ & $10.82 \pm 0.05$ & $10.41 \pm 0.06$ \\
GN-IRS-12 & 2.009 & 12.8 & 79 $\pm$ 15 & $0.006 \pm 0.001$ & $9.31 \pm 0.11$ & $<8.74$ & $10.85 \pm 0.15$ & \nodata \\
GN-IRS-15 & 1.962 & 12.9 & 22 $\pm$ 2 & $0.078 \pm 0.005$ & $9.67 \pm 0.11$ & $<8.84$ & $10.98 \pm 0.05$ & $<10.63$ \\
GN-IRS-18 & 1.799 & 12.5 & 32 $\pm$ 3 & $0.050 \pm 0.001$ & $9.56 \pm 0.10$ & $8.44 \pm 0.09$ & $10.84 \pm 0.05$ & $10.37 \pm 0.10$ \\
GN-IRS-19 & 1.886 & 12.4 & 63 $\pm$ 30 & $0.010 \pm 0.001$ & $9.20 \pm 0.11$ & $8.54 \pm 0.08$ & $10.48 \pm 0.08$ & $10.19 \pm 0.06$ \\
GN-IRS-21 & 1.974 & 12.8 & 47 $\pm$ 6 & $0.110 \pm 0.003$ & $9.56 \pm 0.11$ & $<9.02$ & $10.83 \pm 0.05$ & $<11.77$ \\
GN-IRS-25 & 1.719 & 12.6 & 2 $\pm$ 1 & $0.073 \pm 0.003$ & $9.53 \pm 0.11$ & $<8.74$ & $11.19 \pm 0.06$ & $10.61 \pm 0.08$ \\
GN-IRS-26 & 1.22 & 12.6 & 2 $\pm$ 2 & $0.104 \pm 0.001$ & $9.78 \pm 0.12$ & $8.53 \pm 0.11$ & $11.02 \pm 0.06$ & $10.03 \pm 0.10$ \\
GN-IRS-27 & 0.776 & 12.1 & 35 $\pm$ 6 & $0.038 \pm 0.001$ & $8.90 \pm 0.11$ & $<8.15$ & $10.18 \pm 0.03$ & \nodata \\
GN-IRS-32 & 1.864 & 12.4 & 48 $\pm$ 7 & $0.037 \pm 0.002$ & $9.27 \pm 0.10$ & $<8.24$ & $10.75 \pm 0.05$ & $<10.22$ \\
GN-IRS-38 & 1.007 & 12.0 & 7 $\pm$ 5 & $0.043 \pm 0.001$ & $9.10 \pm 0.11$ & $<8.52$ & $10.38 \pm 0.05$ & $10.24 \pm 0.06$ \\
GN-IRS-42 & 0.971 & 12.1 & 18 $\pm$ 8 & $0.054 \pm 0.001$ & $9.07 \pm 0.11$ & $<8.32$ & $10.53 \pm 0.08$ & $10.31 \pm 0.09$ \\
GN-IRS-48 & 0.827 & 11.8 & 0 $\pm$ 5 & $0.046 \pm 0.001$ & $8.86 \pm 0.11$ & $8.15 \pm 0.13$ & $<10.26$ & \nodata \\
GN-IRS-50 & 0.822 & 11.6 & 0 $\pm$ 5 & $0.043 \pm 0.001$ & $8.75 \pm 0.11$ & $8.07 \pm 0.11$ & $10.13 \pm 0.09$ & \nodata \\
GN-IRS-54 & 0.831 & 11.4 & 34 $\pm$ 2 & $0.015 \pm 0.001$ & $8.19 \pm 0.12$ & $<8.04$ & $<9.92$ & \nodata \\
GN-IRS-55 & 1.99 & 12.4 & 67 $\pm$ 4 & \nodata & $<8.49$ & $<8.36$ & $10.63 \pm 0.09$ & \nodata \\
GN-IRS-57 & 1.96 & 11.8 & 90 $\pm$ 9 & \nodata & $<8.21$ & $<8.33$ & $<10.31$ & \nodata \\
GN-IRS-58 & 1.999 & 12.3 & 32 $\pm$ 2 & $0.072 \pm 0.004$ & $9.12 \pm 0.10$ & $<8.26$ & $<10.55$ & \nodata \\
GN-IRS-61 & 0.842 & 11.7 & 0 $\pm$ 5 & $0.072 \pm 0.001$ & $8.70 \pm 0.11$ & $8.15 \pm 0.08$ & $<10.11$ & \nodata \\
GN-IRS-63 & 1.466 & 12.1 & 9 $\pm$ 1 & $0.092 \pm 0.001$ & $9.25 \pm 0.11$ & $<8.22$ & $10.61 \pm 0.08$ & $10.09 \pm 0.06$ \\
FLS-IRS-289 & 1.844 & 13.1 & 37 $\pm$ 3 & $0.090 \pm 0.002$ & $9.84 \pm 0.10$ & $8.78 \pm 0.14$ & $11.29 \pm 0.02$ & $10.68 \pm 0.07$ \\
FLS-IRS-509 & 2.177 & 12.4 & 77 $\pm$ 2 & \nodata & $9.42 \pm 0.16$ & $<9.33$ & $11.19 \pm 0.09$ & $<11.84$ \\
FLS-IRS-521 & 2.017 & 13.0 & 40 $\pm$ 8 & $0.038 \pm 0.001$ & $9.89 \pm 0.10$ & $<8.81$ & $11.14 \pm 0.06$ & $<11.00$ \\
FLS-IRS-8040 & 0.76 & 12.5 & 38 $\pm$ 3 & $0.038 \pm 0.001$ & $<10.10$ & $<9.42$ & $10.56 \pm 0.03$ & $10.49 \pm 0.05$ \\
FLS-IRS-8226 & 2.119 & 12.8 & 69 $\pm$ 3 & $0.005 \pm 0.001$ & $9.32 \pm 0.13$ & $8.86 \pm 0.10$ & $11.29 \pm 0.04$ & $<10.86$ \\
FLS-IRS-8242 & 2.46 & 13.2 & 94 $\pm$ 3 & $0.023 \pm 0.001$ & $10.36 \pm 0.15$ & $<9.89$ & $11.17 \pm 0.09$ & \nodata \\
FLS-IRS-8493 & 0.647 & 11.6 & 35 $\pm$ 3 & $0.029 \pm 0.001$ & $8.88 \pm 0.31$ & $8.53 \pm 0.30$ & $9.95 \pm 0.06$ & $<9.82$ \\
FLS-IRS-16122 & 1.845 & 12.8 & 88 $\pm$ 9 & $0.016 \pm 0.001$ & $9.34 \pm 0.19$ & $<9.72$ & $<11.22$ & $<10.77$ \\
FLS-IRS-22530 & 1.963 & 13.1 & 21 $\pm$ 5 & $0.077 \pm 0.002$ & $9.89 \pm 0.10$ & $<8.93$ & $11.16 \pm 0.05$ & $<10.37$ \\
FLS-IRS-22722 & 1.802 & 12.6 & 77 $\pm$ 8 & $0.024 \pm 0.001$ & $9.26 \pm 0.15$ & $<9.57$ & $<10.72$ & $<10.73$ \\
\enddata
\tablecomments{By column number: (1) Galaxy identifier. (2) Spectroscopic redshift from the MIRI LRS spectrum. (3) Infrared luminosity, accurate to 0.2 dex. (4) Mid-infrared AGN fraction. (5) 3.3\,$\mu$m PAH EW measured directly from the spectrum. (6)--(9) Extinction-corrected PAH luminosities from the spectral decomposition (Section 4.2), with 1$\sigma$ uncertainties. Uncertainties on $L_{3.3}$ include an additional 0.1\,dex, added in quadrature, for the modelling of the 3.05\,$\mu$m ice and 3.4\,$\mu$m aliphatic features (Section 4.2). Upper limits are 3$\sigma$ and are given where the feature is detected at $<3\sigma$ in the observed (uncorrected) flux. $L_{7.7}$ is the sum of the 7.60 and 7.85\,$\mu$m components and $L_{11.3}$ the sum of the 11.23 and 11.33\,$\mu$m components. \nodata\ indicates that the feature is not covered by the spectrum.}
\end{deluxetable*}

\bibliography{references}
\bibliographystyle{aasjournal}

\end{document}